\documentclass{article}

\usepackage{amsmath,amsfonts}
\usepackage{array}
\usepackage{textcomp}
\usepackage{stfloats}
\usepackage{url}
\usepackage{verbatim}
\usepackage{graphicx}
\usepackage{cite}
\usepackage{nicefrac, xfrac}
 \usepackage{adjustbox}
\usepackage{caption}
\usepackage{subcaption}
\usepackage{booktabs}
\usepackage{float}
\usepackage{multicol}
\usepackage{multirow}
\usepackage{amssymb}
\usepackage[ruled,linesnumbered]{algorithm2e}
\usepackage{hyperref}
\usepackage{url}
\hypersetup{
    colorlinks,
    linkcolor={black!50!black},
    citecolor={black!50!black},
    urlcolor={black!80!black}
}

\usepackage{orcidlink}

\newcommand{\R}{\ensuremath{\mathbb{R}}}
\newcommand{\C}{\ensuremath{\mathbb{C}}}

\newcommand{\s}{\mathbf{s}}
\newcommand{\dd}{\mathbf{d}}
\newcommand{\p}{\mathbf{p}}
\newcommand{\PPhi}{\boldsymbol{\Phi}}

\newcommand{\vv}{\mathbf{v}}

\newcommand{\A}{\mathbf{A}}
\newcommand{\I}{\mathbf{I}}
\newcommand{\n}{\mathbf{w}}
\newcommand{\y}{\mathbf{y}}
\newcommand{\x}{\mathbf{r}}

\newcommand{\mycomment}[1]{}

\usepackage{arxiv}
\usepackage[utf8]{inputenc} 
\usepackage[T1]{fontenc}   
\usepackage{hyperref}       
\usepackage{url}           
\usepackage{booktabs}      
\usepackage{amsfonts}       
\usepackage{nicefrac}       
\usepackage{microtype}      
\usepackage{lipsum}
\usepackage{graphicx}
\graphicspath{ {./images/} }

\newcommand\blfootnote[1]{%
  \begingroup
  \renewcommand\thefootnote{}\footnote{#1}%
  \addtocounter{footnote}{-1}%
  \endgroup
}

\title{Plug-and-Play Regularization on Magnitude with Deep Priors for 3D Near-Field MIMO Imaging}

\author{ \normalfont Okyanus~Oral\,\orcidlink{0000-0001-5059-4351},~\textit{Graduate~Student~Member}, and Figen~S.~Oktem\,\orcidlink{0000-0002-7882-5120},~\textit{Member,~IEEE}}

\begin{document}

\blfootnote{This work is supported by the Scientific and Technological Research Council of Turkey  (TUBITAK) under grant 120E505.}
\blfootnote{O.~Oral, and F.~S.~Oktem are with the Department of Electrical Engineering, METU, Cankaya, Ankara 06800, Turkey (e-mail: ookyanus@metu.edu.tr, figeno@metu.edu.tr).}
\blfootnote{This work has been submitted to the IEEE for possible publication. Copyright may be transferred without notice, after which this version may no longer be accessible.}

\maketitle

\begin{abstract}
Near-field radar imaging systems are used in a wide range of applications such as concealed weapon detection and medical diagnosis. In this paper, we consider the problem of reconstructing the three-dimensional (3D) complex-valued reflectivity distribution of the near-field scene by enforcing regularization on its magnitude. We solve this inverse problem by using the alternating direction method of multipliers (ADMM) framework. For this, we provide a general expression for the proximal mapping associated with such regularization functionals. This equivalently corresponds to the solution of a complex-valued denoising problem which involves regularization on the magnitude. By utilizing this expression, we develop a novel and efficient plug-and-play (PnP) reconstruction method that consists of simple update steps. Due to the success of data-adaptive deep priors in imaging, we also train a 3D deep denoiser to exploit within the developed PnP framework. The effectiveness of the developed approach is demonstrated for multiple-input multiple-output (MIMO) imaging under various compressive and noisy observation scenarios using both simulated and experimental data. The performance is also compared with the commonly used direct inversion and sparsity-based reconstruction approaches. The results demonstrate that the developed technique not only provides state-of-the-art performance for 3D real-world targets, but also enables fast computation. Our approach provides a unified general framework to effectively handle arbitrary regularization on the magnitude of a complex-valued unknown and is equally applicable to other radar image formation problems (including SAR). 
\end{abstract}


\section{Introduction}

Near-field radar imaging systems are recently used in a wide range of applications such as medical diagnosis, through-wall imaging, concealed weapon detection, and nondestructive evaluation~\cite{fioranelli2014frequency,ahmed2012advanced,zhuge2010sparse,anadol2018uwb}.
For various high-resolution imaging applications, there has been a growing interest in using multiple-input multiple-output (MIMO) arrays (i.e. multistatic arrays) that contain spatially distributed transmit and receive antennas~\cite{ahmed2012advanced,zhuge2010sparse,zhuge2012study,ahmed2009near,yanik2019near, anadol2018uwb, kocamis2017optimal}. MIMO arrays offer reduced hardware complexity, cost, and acquisition time compared to the conventional monostatic planar arrays (with colocated transmitter and receiver antennas).

In near-field radar imaging, the three-dimensional (3D) complex-valued scene reflectivity has to be reconstructed from the radar data that is generally acquired using sparse arrays. This requires solving an ill-posed inverse problem. Consequently, the imaging performance greatly depends on the underlying image reconstruction method and the utilization of priors.

Traditional direct inversion methods do not utilize any prior information and are solely derived to obtain a direct solution based on the forward (observation) model expression. These methods generally involve back-projecting measurements to the image domain and then employing a filter-like operation \cite{Zhuge2012RMA,alvarez2016fourier,Zhuge2010Kirchhoff,Marks2017Fourier}. Kirchhoff migration \cite{Zhuge2010Kirchhoff}, back-projection \cite{anadol2018uwb}, and range migration \cite{Zhuge2012RMA} are commonly used direct inversion methods for near-field radar imaging. Although these methods offer low computational complexity, their reconstruction performance substantially degrades in ill-posed settings with limited and noisy data.

Regularization-based methods can yield more successful reconstructions than these traditional methods by incorporating prior information about the unknown 3D image cube into the reconstruction process. One way to utilize prior information is to minimize an appropriately formulated cost function using hand-crafted regularization terms~\cite{hansen2010discrete,oktem2017book,guven2016augmented,miran2021sparse,oktem2019sparsity,Yang2012Near}. Examples include total-variation (TV) \cite{beck2009fast} and $\ell_1$ regularization. These commonly used sparsity priors are motivated by the compressed sensing theory~\cite{Candes2008Intro2CS} and are shown to offer promising imaging performance at various compressive imaging settings including radar imaging~\cite{potter2010,oktem2019sparsity,miran2021sparse,cetin2001feature, Li2015NFCsensing, guven2016augmented, Yang2012Near}. For near-field radar imaging, existing regularized reconstruction methods generally enforce smoothness or sparsity on the complex-valued reflectivity distribution~\cite{Yang2012Near, Wang20223DSparse, oktem2019sparsity,Li2015NFCsensing}. These methods are therefore built on the assumption that the scene reflectivity has locally correlated phase and magnitude. However, for many applications, the phase of the reflectivity at a particular point can be more accurately modeled as random and uncorrelated with the phase at other points~\cite{Munson1984offset,cetin2001feature, potter2010}. This is because phase shift can occur when imaging rough surfaces and also at the air/target interface due to the electrical properties of materials~\cite{Munson1984offset}.
It has been observed in various related SAR works that enforcing regularization only on the magnitude improves the performance compared to enforcing it directly on the complex-valued reflectivity~\cite{cetin2001feature,Cetin2021PnP, guven2016augmented, potter2010}.

With the recent advancements in deep learning, learned reconstruction methods have emerged as powerful alternatives to the regularization-based methods with hand-crafted analytical priors~\cite{ongie2020,monga2020unrolling,Kamilov2023PhysicsBased,Aggarwal2018MODL,zhang2021plug,Cetin2021PnP}. 
These methods can utilize deep neural networks (DNNs) to learn data-driven denoiser priors and then incorporate these priors in a model-based reconstruction as a regularizer. The main motivation behind exploiting DNN-based denoisers in these approaches lies in the observation that DNNs provide state-of-the-art performance in denoising~\cite{ongie2020}.
Learned Plug-and-Play (PnP) regularization \cite{Venkatakrishnan2013PnP,Cetin2021PnP,zhang2021plug,Kamilov2023PhysicsBased, Chan2017PnPADMM} 
and unrolling-based methods \cite{gregor2010learning,Aggarwal2018MODL,monga2020unrolling}
are examples of such approaches. In particular, the key idea in learned PnP methods is to first learn a deep denoiser prior from training data and then 
substitute this denoiser in place of proximal operator in the used optimization framework. Commonly used frameworks for this purpose include alternating direction method of multipliers (ADMM)~\cite{boyd2011admm} and proximal gradient descent~\cite{beck2009fast}. Another approach for exploiting deep priors is based on unrolling, which converts an iterative method that utilize deep-priors, such as PnP, into an end-to-end trainable network~\cite{monga2020unrolling, Aggarwal2018MODL,ongie2020}. Although both learned PnP- and unrolling-based approaches yield state-of-the-art reconstruction quality, PnP methods have the advantage of adaptability to different imaging settings and significantly less training time.

Despite the recent success of PnP methods with deep priors, most of these approaches have been developed for 2D or real-valued image reconstruction problems~\cite{ongie2020, Kamilov2023PhysicsBased, zhang2021plug, Venkatakrishnan2013PnP, Cetin2021PnP}. Furthermore, there is no study on such methods for near-field radar imaging where we encounter a 3D complex-valued image reconstruction problem. 

In this paper, we develop a novel and efficient PnP method for reconstructing the 3D complex-valued reflectivity distribution of the near-field scene from sparse MIMO measurements. Due to the random phase nature of the scene reflectivities in various applications, we formulate the image formation problem by exploiting regularization on the magnitude of the reflectivity function. We provide a general expression for the proximal mapping associated with such regularization functionals operating on the magnitude. By utilizing this expression, we develop a computationally efficient PnP reconstruction method that consists of simple update steps. To utilize within the developed PnP framework, we also train a 3D deep denoiser that can jointly exploit range and cross-range correlations. The source codes of this developed approach are available at \url{https://github.com/METU-SPACE-Lab/PnP-Regularization-on-Magnitude}.

Our approach provides a unified PnP framework to effectively handle arbitrary regularization on the magnitude of a complex-valued unknown, which appears to be missing in the previous related radar imaging works~\cite{Cetin2021PnP,guven2016augmented}. The effectiveness of the developed learning-based PnP approach is illustrated in microwave imaging  under various compressive and noisy observation scenarios using both simulated data and experimental measurements. We also compare the performance with the commonly used traditional methods (back-projection and Kirchhoff migration), and with the sparsity-based approaches involving $\ell_1$ and TV regularization. 

Compared to the earlier works in near-field MIMO radar imaging, the developed technique not only provides state-of-the-art reconstruction performance for 3D real-world targets, but also enables fast computation. In particular, compared to the traditional direct inversion methods and sparsity-based approaches, the developed reconstruction technique achieves the best reconstruction quality at compressive settings with both simulated and experimental data. Some preliminary results of this research have been presented in \cite{oral2023plug}. Here, we provide a more complete treatment of the theoretical work, and illustrate the performance through extensive simulations as well as using real-world experimental measurements. Different than the related learning-based works in near-field MIMO radar imaging~\cite{Cheng2020Compressive, smith2022vision, manisali2024efficient,wei2022learning}, our approach is a deep prior-based PnP approach developed for imaging 3D extended targets. In particular, the works in \cite{Cheng2020Compressive, smith2022vision,manisali2024efficient} present deep learning-based non-iterative reconstruction methods by refining an initial analytical reconstruction using DNNs. Other learning-based work in \cite{wei2022learning} develops an unrolling-based method. But unlike our approach, this method is not DNN-based (i.e. not deep prior-based) and only learns the hyperparameters (such as soft threshold and regularization parameters) of the unrolled $\ell_1$ regularization-based reconstruction algorithm.

To the best of our knowledge, our approach is the first deep prior-based PnP approach developed for near-field radar imaging where we encounter a 3D complex-valued image reconstruction problem. A related PnP work in SAR imaging~\cite{Wang20223DSparse} utilizes 2D analytical (but not deep) denoising priors to reconstruct 3D extended targets. This approach also considers regularization on the complex-valued reflectivity. Differently, our approach exploits regularization on the magnitude of the reflectivity due to its random phase nature in various applications. There is also a related learned PnP approach with magnitude regularization which has been developed for 2D (far-field) SAR imaging \cite{Cetin2021PnP}. However, this method requires an inefficient iterative computation to update the phase. In contrast, our approach does not have a phase update step and all the other update steps are simple and efficient to compute thanks to the closed-form expression used for the proximal mapping. The presented closed-form expression for the proximal mapping associated with arbitrary regularization on the magnitude also provides a generalization of the proximal mappings associated with TV and $\ell_1$ regularization on magnitude \cite{guven2016augmented}. Hence our PnP framework provides a generalizable and powerful means for effectively enforcing arbitrary regularization on magnitude, and is equally applicable to other radar image formation problems (including SAR).

The main contributions of this paper can be summarized as follows:
\begin{itemize}
    \item Providing a unified PnP framework to effectively handle arbitrary regularization on the magnitude of a complex-valued unknown (involving random phase),
    \item Development of a novel deep learning-based plug-and-play reconstruction method for 3D complex-valued imaging with application to near-field MIMO radar imaging,
    \item Comprehensive experiments on synthetic 3D scenes with quantitative and qualitative analysis by considering various compressive and noisy
    observation scenarios, 
    \item Performance evaluation with experimental measurements to demonstrate reconstruction of 3D real-world targets, and comparison with the commonly used direct inversion and regularized reconstruction methods.
\end{itemize}

The paper is organized as follows. In Section \ref{Section:ObservationModel} we describe the working principle of a near-field MIMO radar imaging system and introduce the observation model. In Section \ref{Section:PNPMethod} we formulate the inverse problem by enforcing regularization on the magnitude and then develop our plug-and-play approach. The architecture of the deep denoiser utilized for learned PnP reconstruction is also presented here. Section \ref{Section:Results} presents the imaging results for various compressive and noisy observation scenarios. The details of the simulated and experimental settings considered, and the training procedure are also presented here. We conclude the paper by providing final remarks in Section \ref{Section:Conclusion}.

\section{Observation Model}
\label{Section:ObservationModel}
\begin{figure}[H]
    \centering
    \includegraphics[width=8cm]{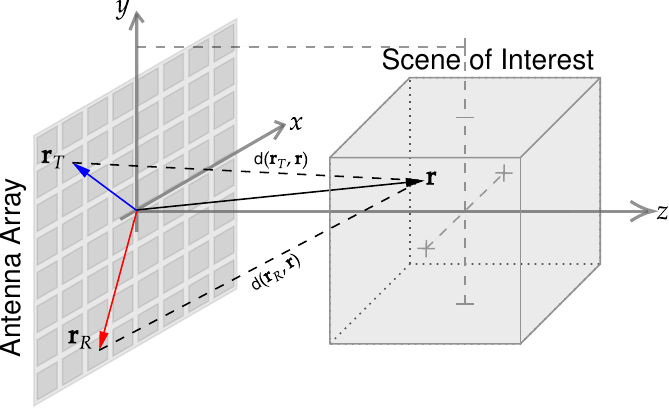}
    \caption{Schematic view of a near-field MIMO radar imaging system.}
    \label{fig:GeneralMIMOArray}
\end{figure}

In this section, we present the image formation model that relates the near-field MIMO array measurements to the reflectivity distribution of the scene. Consider the general MIMO imaging setting illustrated in Fig.~\ref{fig:GeneralMIMOArray} with spatially distributed transmit and receive antennas on the antenna array located at $z=0$. In order to infer the 3D reflectivity distribution of the scene, each transmit antenna, located at $\x_T=[x_T,y_T,0]^T$, illuminates the scene with a pulse signal and the scattered field from the scene is measured by a receive antenna, located at $\x_R=[x_R,y_R,0]^T$.

Under Born approximation, time-domain response of a single point-scatterer with reflectivity $s(\x)$ and located at $\x=[x,y,z]^T$ can be expressed as follows~\cite{Zhuge2012RMA}: 
\begin{equation}
 \tilde{y}(\x_T,\x_R,t)=\frac{p(t-\frac{d(\x_T,\x)}{c}-\frac{d(\x_R,\x)}{c})}{4\pi\; d(\x_T,\x)\; d(\x_R,\x)}s(\x).
\end{equation}
Here $\tilde{y}(\x_T,\x_R,t)$ denotes the time-domain measurement acquired using the transmit and receive antenna pair located respectively at $\x_T$ and $\x_R$ due to a single scatterer. The transmitted pulse is denoted by $p(t)$, and $c$ denotes the speed of light. The distances of the scatterer to the corresponding transmitter and receiver are given by $d(\x_T,\x)=\|\x_T-\x\|_2$ and $d(\x_R,\x)=\|\x_R-\x\|_2$ respectively. 

By taking 1D Fourier transform over $t$, the received signal due to a single scatterer can be expressed in the temporal frequency domain as follows:
\begin{align}
     \tilde{y}(\x_T,\x_R,k) = h(\x_T,\x_R,k,\x) s(\x),
\label{eqn:MeasSinglePTScatterer}
\end{align}
where
\begin{equation}
h(\x_T,\x_R,k,\x)=p(k)\frac{e^{-jk\left(d(\x_T,\x)+d(\x_R,\x)\right)}}{4\pi\;d(\x_T,\x)\; d(\x_R,\x)},
\label{eqn:kernel}
\end{equation}
and $k= \frac{2\pi}{c} f$ denotes the frequency-wavenumber whereas $f$ denotes the temporal frequency. Using \eqref{eqn:MeasSinglePTScatterer}, the measurement, $y(\x_T,\x_R,k)$, due to an extended target can be expressed as the superposition of these responses from point-scatterers:
\begin{align}
      y(\x_T,\x_R,k) &= \underset{x}{\int}\underset{y}{\int}\underset{z}{\int} h(\x_T,\x_R,k,\x)s(\x) \; d\x.
      \label{eqn:CTMeasurements}
\end{align}

Since the measurements are discrete, and the image reconstruction algorithm will be run on a computer, a discrete forward model is needed. For this, the coordinate variables are discretized based on the expected range and cross-range resolutions of the used MIMO imaging system~\cite{Zhuge2012RMA}. Then the discretized scene reflectivity values can be related to the discrete measurements obtained using different transmitter-receiver pairs and frequency steps as 
\begin{align}
y(\x_{T_m},\x_{R_m},k_m)&=\sum_n h(\x_{T_m},\x_{R_m},k_m,\x_n) s(\x_n).
\label{eqn:DTMeasurements}
\end{align}
Here the subscript $m$ indicates the location of the transmitting and receiving antennas as well as the frequency used in the $m$th measurement. Moreover, the subscript $n$ indicates the voxel number in the discretized 3D scene. 

By using lexicographical ordering, the measurements and the reflectivity values of the image voxels are put into the following vectors:
\begin{align}
    \y&=[y(\x_{T_1},\x_{R_1},k_1),\dots,y(\x_{T_M},\x_{R_M},k_M)]^T\in\C^M,\\
    \s&=[s(\x_1),\dots,s(\x_N)]^T\in \C^N,
\end{align}
where $M$ and $N$ respectively represent the number of  measurements and voxels. Then using \eqref{eqn:DTMeasurements} we can express the noisy measurements in matrix-vector form as follows:
\begin{equation}
    \y = \A\s + \n.
    \label{eqn:forward-model-MVprod}
\end{equation}
The matrix $\A \in \C^{M\times N}$ is the observation matrix whose $(m, n)$th element is given by 
\begin{align}
    \A_{m,n}&=h(\x_{T_m},\x_{R_m},k_m,\x_n),
\end{align}
which represents the contribution of the $n$th voxel at location $\x_n$ to the $m$th measurement taken using the transmitter at $\x_{T_m}$, receiver at $\x_{R_m}$, and frequency $ \frac{c}{2\pi}k_m$. Also 
$\n\in \C^M$ represents the additive noise vector. We assume white Gaussian noise since it commonly holds in practical applications of interest. Hence each noise component is uncorrelated over different voxels and has variance $\sigma_w^2$. 

\section{Plug-And-Play Reconstruction Approach} 
\label{Section:PNPMethod}
In this section, we first formulate the inverse problem by enforcing regularization on the magnitude and then develop our plug-and-play approach using the ADMM framework. The architecture of the 3D deep denoiser utilized for learned PnP reconstruction is also presented here.

\subsection{Inverse Problem}
In the inverse problem, the goal is to estimate the 3D complex-valued reflectivity field, $\s$, from the acquired radar measurements, $\y$. This corresponds to solving an under-determined problem with sparse measurements $M\ll N$. As a result, the reconstruction quality greatly depends on the utilization of priors. A systematic approach to regularization is to incorporate the prior knowledge about unknown solution in a deterministic or stochastic setting, and leads to a minimization with a regularization functional penalizing the solutions that do not comply with the assumed prior information~\cite{hansen2010discrete, oktem2017book}. 

Due to the random phase nature of the unknown scene reflectivities, we formulate the inverse problem using a regularization functional, $\mathcal{R}(|\cdot|)$, that only operates on the magnitude:
\begin{equation}
    \underset{s}{\min} \;\mathcal{R}(|\s|) \text{ subject to } \|\y-\A\s\|_2 \leq \epsilon
    \label{eqn:inverse-problem}    
\end{equation}
where $\epsilon$ is a parameter that should be chosen based on the noise variance (i.e. $\sqrt{M\cdot\sigma^2_w}$), and $|\s|$ denotes the magnitude of the reflectivity vector $\s$.

\subsection{Variable Splitting and ADMM}
To solve this regularized inverse problem, we first convert the constrained problem in \eqref{eqn:inverse-problem} to an unconstrained one using the penalty function, $\iota_ {\|\y-\vv_1\|_2\leq \epsilon}(.)$, and then apply variable splitting as follows:
\begin{align}
\label{eqn:inverse-problem-CSALSA}
& \underset{s, \vv_1,\vv_2}{\min} \left( \iota_ {\|\y-\vv_1\|_2\leq \epsilon}(\vv_1) + \mathcal{R}(|\vv_2|) \right)\\
&\text{subject to } \A\s-\vv_1=0 \;,\; \s-\vv_2=0  \nonumber
\end{align}
Here the indicator function $\iota_ {\|\y-\vv_1\|_2\leq \epsilon}(\vv_1)$ takes value $0$ if the constraint in \eqref{eqn:inverse-problem} is satisfied and $+\infty$ otherwise, whereas $\vv_1$, $\vv_2$ are the auxiliary variables. 

We solve the optimization problem in \eqref{eqn:inverse-problem-CSALSA} with the C-SALSA approach \cite{afonso2010csalsa}. In the corresponding ADMM framework~\cite{boyd2011admm}, we first obtain the associated augmented Lagrangian form given by 
\begin{align}
    \nonumber
    \mathcal{L}_{\rho_1,\rho_2}&(\s,\vv_1,\vv_2,\dd_1,\dd_2) =\\
    \nonumber
    +\iota&_{\|\y  -\vv_1\|_2 \leq \epsilon}(\vv_1 )+ \frac{\rho_1}{2}\| \A\s-\vv_1-\dd_1 \|_2^2 - \frac{\rho_1}{2} \|\dd_1 \|^2_2\\
    &+\mathcal{R}(|\vv_2| ) + \frac{\rho_2}{2} \|\s-\vv_2-\dd_2 \|_2^2 - \frac{\rho_2}{2} \|\dd_2\|_2^2  
\end{align}
Here $\dd_1$, $\dd_2$ denote the dual variables for $\A\s$ and $\s$, and $\rho_1,\rho_2 \in \R^+$ are the penalty parameters for the auxiliary variables $\vv_1$ and $\vv_2$. We then alternatively minimize this augmented Lagrangian function over $\s$, $\vv_1$, and $\vv_2$ to obtain the update steps for these variables. 

Firstly, the minimization over $\s$ corresponds to solving a least-squares problem with the following normal equation:
\begin{equation}
    (\A^H\A + \kappa \I)\s^{l+1} =\A^H(\vv^l_1+\dd^l_1) + \kappa(\vv^l_2 + \dd^l_2) \\
\label{eqn:x-update}
\end{equation}
where the superscript $l$ is the iteration count, and $\kappa\triangleq\frac{\rho_1}{\rho_2}$ is a hyper-parameter that needs to be adjusted. 
Since solving this normal equation using matrix inversion is impractical due to the large size, we instead use few conjugate-gradient (CG) iterations to update the scene reflectivity $\s$. 

Secondly, the minimization over $\vv_1$ corresponds to the proximal operator of the penalty function $\iota_{\|\y  -\vv_1\|_2 \leq \epsilon}(\cdot)$, which can be computed as the projection of $\A\s^{l+1}-\dd^l_1$ onto $\epsilon$-radius hyper-sphere with center $\y$ as follows:
\begin{align}    
    \vv_1^{l+1} =\y+\begin{cases}\epsilon\frac{\A\s^{l+1}-\dd^l_1-\y}{\|\A\s^{l+1}-\dd^l_1-\y\|_2},& \text{\hspace{-5pt}if } \|\A\s^{l+1}-\dd^l_1 -\y\|_2>\epsilon \\
     \A\s^{l+1}-\dd^l_1-\y,& \text{\hspace{-5pt}if } \|\A\s^{l+1}-\dd^l_1-\y\|_2 \leq \epsilon \\
     \end{cases}
     \label{eqn:proj}
\end{align}

Lastly, the minimization over $\vv_2$ corresponds to the proximal operator for the regularization function, $\mathcal{R}(|\cdot|)$, that operates on the magnitude of the complex-valued vector $\vv_2$:
\begin{equation}
    \vv^{l+1}_2=\mathbf{\Psi}_{\alpha \mathcal{R(|\cdot|)}}(\s^{l+1}-\dd_2^l)
    \label{eqn:update3}
\end{equation}
where $\mathbf{\Psi}_{\alpha \mathcal{R(|\cdot|)}}$ is the respective proximal operator given by
\begin{equation}
 \mathbf{\Psi}_{\alpha \mathcal{R(|\cdot|)}}(\p) \triangleq \arg \underset{\vv}{\min} \left(  \alpha \mathcal{R}(|\vv|) + \frac{1}{2}\|\vv-\p\|^2_2 \right)\\
    \label{eqn:prox-mag}
\end{equation}
for a complex-valued vector $\p$, with $\alpha \triangleq \frac{1}{\rho_2}$ determining the amount of regularization. This update step corresponds to solving a denoising problem for a complex-valued unknown, $\vv$, with regularization enforced on its magnitude and noisy observation given as $\p$. To develop a computationally efficient PnP reconstruction method that consists of simple update steps, we provide a general expression for the solution of this denoising problem (equivalently, for the proximal operator in \eqref{eqn:prox-mag}). This will enable us to effectively handle arbitrary regularization on the magnitude, which appears to be missing in the previous radar imaging works.

\subsection{Denoising with Regularization on Magnitude} 

\begin{figure*}[t]
    \centering
    \includegraphics[width=\textwidth]{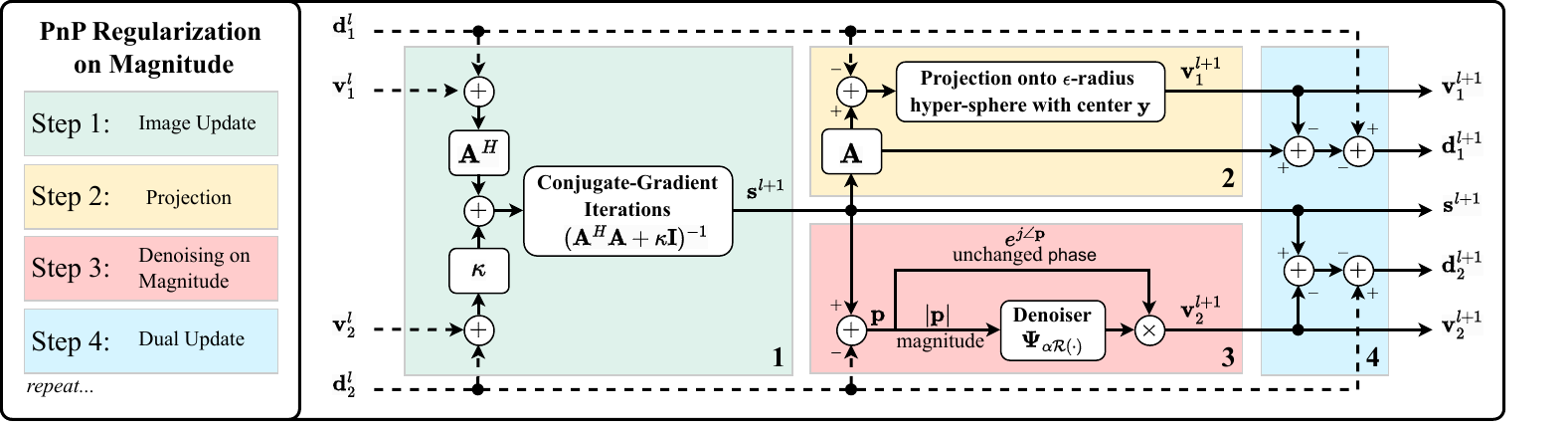}
    \caption{Developed PnP Method for Complex-valued Reconstruction with Regularization on Magnitude.}
    \label{fig:BlockDiagram}
\end{figure*}

In this section, we provide a general expression for the solution of the complex-valued denoising problem in \eqref{eqn:prox-mag} which involves regularization on the magnitude. For this, we first express each complex-valued vector as a product of a diagonal phase matrix and a magnitude vector as follows:
 \begin{equation}
 \vv=\PPhi_\vv|\vv|,\quad \p=\PPhi_\p|\p|, 
 \end{equation}
where $\PPhi_\vv = \text{diag}(e^{j\angle\vv})$ and $\PPhi_\p=\text{diag}(e^{j\angle\p})$ are complex-valued unitary matrices that contain the phase of the vectors $\vv$ and $\p$ on their diagonals, respectively, whereas $|\vv|$ and $|\p|$ represent real-valued and non-negative vectors that contain the respective magnitudes. By using these expressions, the optimization problem in \eqref{eqn:prox-mag} can be viewed as a joint minimization over the magnitude and phase of $\vv$: 
\begin{equation}
\underset{|\vv|, \,\angle\vv}{\min} \left(  \alpha \mathcal{R}(|\vv|) + \frac{1}{2}\|\PPhi_\vv|\vv|-\PPhi_\p|\p|\|^2_2 \right)
\label{eqn:equivalent optimization}
\end{equation}

This joint minimization problem is equivalent to
\begin{subequations}
\begin{align}
\label{eqn:split-opt-1}
&\underset{|\vv|}{\min} \left( \underset{\angle\vv}{\min} \left(\alpha \mathcal{R}(|\vv|) + \frac{1}{2}\|\PPhi_\vv|\vv|-\PPhi_\p|\p|\|^2_2 \right)\right)\\
\label{eqn:split-opt-2}
\equiv \;&\underset{|\vv|}{\min} \left( \alpha \mathcal{R}(|\vv|) +  \underset{\angle\vv}{\min} \left(\frac{1}{2}\|\PPhi_\vv|\vv|-\PPhi_\p|\p|\|^2_2\right)\right)    
\end{align}
\end{subequations}
Hence to solve this complex-valued denoising problem, our strategy is to first solve the minimization over the phase, $\angle\vv$, in closed-form, and then by substituting the optimal phase solution, $\angle\hat{\vv}$, to the above cost function, to solve the remaining minimization over the magnitude, $|\vv|$. 

For minimization over the phase, we have 
\begin{eqnarray}
\angle\hat{\vv} &=& \arg\underset{\angle\vv}{\min} \left(\frac{1}{2}\|\PPhi_\vv|\vv|-\PPhi_\p|\p|\|^2_2\,\right)\label{eqn:phase-opt} \\
&=& \arg\underset{\angle\vv}{\min} \left(\frac{1}{2}\|\PPhi^H_\p\PPhi_\vv|\vv|-|\p|\|^2_2\right)
\label{eqn:phase-opt-3}  
\end{eqnarray}
where the last expression follows from the unitary property of the phase matrices, i.e. $\PPhi^H_\p=\PPhi_\p^{-1}$. After expanding the $\ell_2$ norm expression and simplifying it using the unitary property of phase matrices and omitting the terms that do not depend on the phase $\angle\vv$,
we obtain 
\begin{equation}
\angle\hat{\vv} =  \arg\underset{\angle\vv}{\max} \left(\frac{1}{2}\left(|\vv|^T \PPhi_\p\PPhi_\vv^*|\p| +|\p|^T \PPhi^*_\p\PPhi_\vv |\vv|\right) \right)
\label{eqn:minimize-phase-simplified-2}
\end{equation}
Here we also use the fact that $|\vv|$ and $|\p|$ are real-valued and hence their Hermitian transpose is simply equal their transpose, and since phase matrices are diagonal, their Hermitian transpose is simply equal their conjugation. Using the diagonality of the phase matrices, this further simplifies to
\begin{equation}
 \angle\hat{\vv} =\arg\underset{\angle\vv}{\max} \left( |\p|^T\mathfrak{Re}\{\PPhi^*_\p\PPhi_\vv\}|\vv| \right). 
\label{eqn:minimize-phase-simplified-3}
\end{equation}
Hence to find the optimal phase, we need to maximize $\sum^N_{n=1} |p_n||v_n|\cos(\angle v_n-\angle p_n)$ over all elements $\angle v_n$ of the vector $\angle \vv$. Since each term in this summation contains only one element of $\angle\vv$, maximization can be decoupled for each element, which yields $\angle p_n$ as the optimal value of $\angle v_n$. This shows that the optimal phase, $\angle\hat{\vv}$, for the denoising problem in \eqref{eqn:equivalent optimization} is equal to the phase of the given noisy observation $\p$:
\begin{equation}
    \angle\hat{\vv} = \angle \p.
\end{equation}
That is, the proximal mapping of a function that operates on the magnitude of a complex-valued vector must directly pass the phase values of
the proximal point.

After solving the minimization over the phase in closed-form, we now substitute the optimal phase solution, $\angle\hat{\vv}$, to the cost function in \eqref{eqn:split-opt-2} and consider the remaining minimization over the magnitude, $|\vv|$:
\begin{equation}
     |\hat{\vv}|=  \arg\underset{|\vv|}{\min} \left( \alpha \mathcal{R}(|\vv|) +  \frac{1}{2}\||\vv|-|\p|\|^2_2 \right)
    \label{eqn:prox-mag-treatedmag}
\end{equation}
where we use the unitary property of the phase matrix $\PPhi_\p$ as before. Note that this expression is equivalent to the Moreau proximal mapping, $\mathbf{\Psi}_{\alpha \mathcal{R(\cdot)}}$, associated with the regularization function $\mathcal{R(\cdot)}$ and applied on the magnitude $|\p|$. Hence the optimal magnitude $|\hat{\vv}|$ for the denoising problem in \eqref{eqn:equivalent optimization}
corresponds to denoising of the magnitude of the noisy observation $\p$ with noise variance $\alpha$:
\begin{align}
    |\hat{\vv}|= \mathbf{\Psi}_{\alpha \mathcal{R(\cdot)}}(|\p|).
\end{align}
For the scalar-valued case, a similar derivation is encountered in \cite{Fessler2021Prox}.

Therefore, the solution of the complex-valued denoising problem in \eqref{eqn:prox-mag} with magnitude regularization can be computed as
\begin{equation}
    \mathbf{\Psi}_{\alpha \mathcal{R(|\cdot|)}}(\p) = e^{j\angle\p}\odot\mathbf{\Psi}_{\alpha \mathcal{R(\cdot)}}(|\p|),
    \label{eqn:prox-final}
\end{equation}
where $\odot$ denotes element-wise multiplication. 
This corresponds to denoising the magnitude of $\p$ using the proximal (denoising) operator $\mathbf{\Psi}_{\alpha \mathcal{R(\cdot)}}$ and merging the denoised magnitude with the unprocessed phase of $\p$. Since \eqref{eqn:prox-final} decouples the magnitude and phase solutions, it enables us to use real-valued denoisers (proximal operators) $\mathbf{\Psi}_{\alpha \mathcal{R(\cdot)}}$ for the solution of the complex-valued denoising problem (in~\eqref{eqn:prox-mag}). 

\subsection{Developed PnP Reconstruction Method}

The steps of the developed PnP method are summarized in Algorithm~\ref{alg:ProposedMethod} and illustrated in Fig.~\ref{fig:BlockDiagram}.
Each iteration of the algorithm mainly consists of four computationally efficient update steps. The first step is the update of the image $\s$ as given in line \ref{alg-step:s-update-LS} and carried out using few CG
iterations. The second step is the update of the auxiliary variable $\vv_1$ by computing the projection given in line \ref{alg-step:v1-update-Proj-2} and efficiently computed using scaling operations. The third step is the complex-valued denoising step given in line \ref{alg-step:v2-update-Denoising} to update the auxiliary variable $\vv_2$. As shown, this denoising is equivalent to directly passing the phase but denoising the magnitude of $\s^{l+1}-\dd^l_2$ using the proximal operator $\mathbf{\Psi}_{\alpha \mathcal{R(\cdot)}}$. To exploit data-driven deep priors, we use a trained denoiser as proximal operator, as explained in the next section. The last steps are the dual-updates given in lines \ref{alg-step:dual-update-1} and \ref{alg-step:dual-update-2}.

\begin{algorithm}
   \SetAlgoLined
   \caption{PnP Regularization on Magnitude for Complex-Valued Reconstruction}
    \label{alg:ProposedMethod}

   \SetKw{inp}{inputs:}
   \inp{$\mathbf{\Psi}_{\alpha \mathcal{R(\cdot)}}$, $\y$, $\A$, $\s^0,\vv_2^0,\vv_1^0$, $\epsilon>0$, $\kappa>0$, $\alpha>0$}\\
   $\dd_1^0,\dd_2^0\gets \mathbf{0}$, $l\gets 0$\\
   
   \Repeat{some stopping criterion is satisfied}{
   $\s^{l+1} = (\A^H\A + \kappa \I)^{-1}(\A^H(\vv^l_1+\dd^l_1) + \kappa(\vv^l_2 + \dd^l_2))$\label{alg-step:s-update-LS}\\ 
    
    $\mathbf{u}^l=\A\s^{l+1}-\dd^l_1$\label{alg-step:v1-update-Proj-1}\\
    
    $\vv_1^{l+1}=\y+\begin{cases}\epsilon\frac{\mathbf{u}^l-\y}{\|\mathbf{u}^l-\y\|_2},& \text{\hspace{-5pt}if } 
    \|\mathbf{u}^l -\y\|_2>\epsilon \\
     \mathbf{u}^l-\y,& \text{\hspace{-5pt}if } \|\mathbf{u}^l-\y\|_2 \leq \epsilon \\
     \end{cases}$\label{alg-step:v1-update-Proj-2}\\

    \vspace{0.05in}
    $\vv^{l+1}_2 = e^{j\angle(\s^{l+1}-\dd^l_2)}\odot\mathbf{\Psi}_{\alpha \mathcal{R(\cdot)}}(|\s^{l+1}-\dd^l_2|)$\label{alg-step:v2-update-Denoising}\\

     $\dd^{l+1}_1=\dd^{l}_1-(\A\s^{l+1}-\vv^{l+1}_1)$ \label{alg-step:dual-update-1}\\
     
     $\dd^{l+1}_2=\dd^{l}_2-(\s^{l+1}-\vv^{l+1}_2)$ \label{alg-step:dual-update-2}\\
     
   $l\gets l+1$\\
   }
    \SetKw{kWoutput}{output:}
    \kWoutput $\s^{l}$
\end{algorithm}

Note that our development is implicit about the choice of the regularizer ($\mathcal{R}(|\cdot|)$) and the related proximal operator ($\mathbf{\Psi}_{\alpha \mathcal{R(\cdot)}}$). Therefore, we can efficiently adopt plug-and-play framework, which enables the utilization of powerful priors, such as deep denoisers, in place of the proximal operator, without explicitly specifying the regularizer.

Moreover, our PnP approach provides a generalizable and powerful means for efficiently handling arbitrary regularization on the magnitude of a complex-valued unknown. Our approach is applicable with any forward model matrix $\A$, and hence can be used for other complex-valued image formation problems including SAR reconstruction.

\subsection{3D Deep Denoiser for Learned PnP Reconstruction}
\begin{figure}[]
    \centering
    \includegraphics[width=8cm]{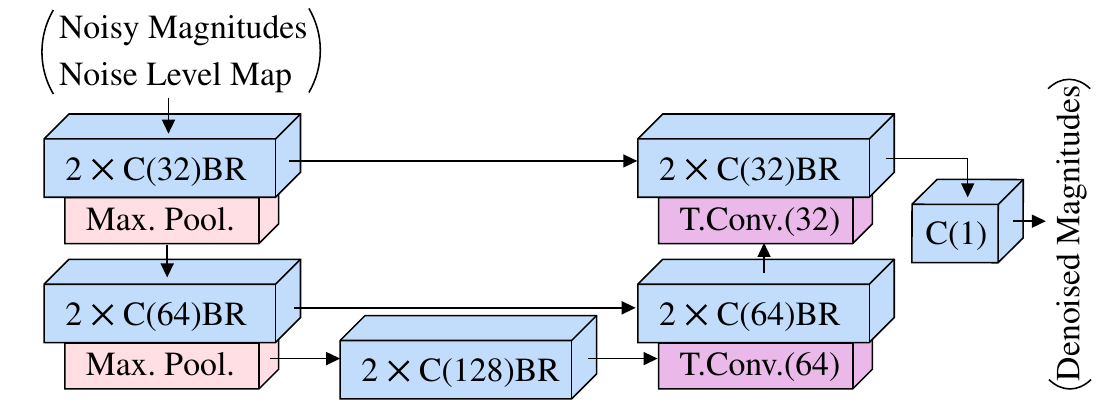}
    \caption{Network architecture of the proposed 3D deep denoiser. “C”, “B”, “R”, “Max. Pool.” and “T.Conv.” represent 3D convolution, batch normalization, ReLU activation, max-pooling operation, and transposed convolution, respectively. The number of output channels is denoted inside parentheses.}
    \label{fig:unet}
\end{figure}

Following the success of convolutional neural networks (CNN) on denoising~\cite{ongie2020,Zhang2017DNCNN,zhang2021plug}, we train and deploy a deep CNN-based denoiser for the third step of our PnP approach. Our denoiser is a 3D U-net developed based on the 2D U-net architecture in \cite{ronneberger2015u} and is shown in Fig.~\ref{fig:unet}. To be able to effectively handle a wide range of noise levels, our denoiser is designed for non-blind Gaussian denoising similar to \cite{zhang2021plug}, and hence takes as input also the noise level. This non-blind denoiser replaces the proximal operator $\mathbf{\Psi}_{\alpha \mathcal{R(\cdot)}}$ in line \ref{alg-step:v2-update-Denoising} of the Algorithm~\ref{alg:ProposedMethod}, which is used to denoise the input magnitudes.

The proposed denoiser is a 3-level encoder-decoder architecture with repeated 3D convolutional blocks (C) followed by batch normalization (B) and ReLU (R). Due to 3D processing, the denoiser can jointly exploit range and cross-range correlations. On each level, max pooling (Max. Pool.) is used to reduce the spatial size of the input tensor by a factor of 2 in each dimension and transposed convolution blocks (T.Conv.) are used to increase by 2. At each decoding level, the output of the transposed convolution block is concatenated with the encoder outputs. The concatenated outputs are then fed to the respective decoding blocks. A single-channel 3D convolution block follows the last decoding block. The number of output channels of all convolutional blocks is indicated inside parentheses in Figure \ref{fig:unet}.

The input of our U-net is the 3D  reflectivity magnitude that will be denoised and the 3D noise level map. The noise level map enables to adjust the amount of denoising in our non-blind denoiser network and its values are set to the constant $\sqrt{\alpha}$ in \eqref{eqn:prox-mag-treatedmag}. The output of the U-net is the 3D denoised reflectivity magnitude.

\section{Experiments and Results}
\label{Section:Results} 

We now demonstrate the effectiveness of the developed learning-based PnP approach under various compressive and noisy observation scenarios in microwave imaging. For this, we first train the implemented denoiser using a synthetically generated large dataset consisting of 3D extended targets. We then perform comprehensive experiments on synthetic 3D scenes, and comparatively evaluate the performance with the widely used back-projection (BP) and Kirchhoff migration (KM) algorithms, as well as using sparsity-based regularization in the form of isotropic total-variation (TV) and $\ell_1$. Lastly, we illustrate the performance with experimental measurements to demonstrate the successful reconstruction of 3D real-world targets.

\subsection{Training of the 3D Deep Denoiser} 
Because a large experimental dataset is not available for microwave imaging, we use a synthetic dataset~\cite{manisali2024efficient} to train our denoiser network. The utilized synthetic dataset consists of randomly generated complex-valued image cubes of size $25 \times25 \times 49$.  
We use 800 image cubes for training, 100 image cubes for testing, and another 100 image cubes for validation. Each synthetic image cube is obtained by randomly generating 15 points within the cube and then applying a 3D Gaussian filter to convert these points to a volumetric object. The magnitudes are normalized (via sigmoid function) to 1, while adding a random phase to each image voxel from a uniform distribution between 0 and $2\pi$.

The denoiser network replaces the proximal operator $\mathbf{\Psi}_{\alpha\mathcal{R}(\cdot)}$ in line \ref{alg-step:v2-update-Denoising} of the Algorithm~\ref{alg:ProposedMethod} with the goal of denoising the reflectivity magnitudes. We accordingly train our deep denoiser by minimizing the mean squared error between the 3D ground truth magnitudes and Gaussian noise added magnitudes on 800 training scenes. At each iteration of training, a new Gaussian noise realization is added to each ground truth magnitude by randomly and uniformly choosing the noise standard deviation, $\sigma_\nu$, from the interval $[0,0.2]$. In addition, the constant noise level map is formed using this value for noise standard deviation, i.e. $\sqrt{\alpha} = \sigma_\nu$, and concatenated to the 3D noisy magnitude. As a result, the network learns to denoise the reflectivity magnitudes in a non-blind manner. 

For training, we use a batch size of 16 with the maximum number of epochs set as $2000$. We utilize Adam optimizer \cite{kingma2014adam} with an initial learning rate of $10^{-3}$, and drop the learning rate by a factor of $10$ if the validation loss does not improve for 25 epochs. We stop the training when the validation loss does not improve for 50 epochs. At the end of training, we use the network weights that provide the minimum validation loss. Training takes approximately 15 minutes on NVIDIA GeForce RTX 3080 Ti GPU using PyTorch 1.12.0 with CUDA Toolkit 11.6.0 in Python 3.10.6. The successful denoising performance of this trained network is demonstrated in the provided supplementary document in comparison with other denoising approaches.

To analyze the performance of our learning-based PnP approach, we use the same trained denoiser without any modification for both simulated and experimental data.

\subsection{Performance Analysis with Simulated Data} 
We first analyze the performance of the developed imaging technique at various noise and compression levels using the synthetic scenes in the test dataset. For this, we consider a microwave imaging setting similar to Fig.~\ref{fig:GeneralMIMOArray}. The scene of interest has physical dimension of $30$ cm $\times$ $30$ cm $\times$ $30$ cm, and its center is located $50$ cm away from the antenna array. 

\begin{figure}[]
    \centering
    \includegraphics[width=0.30\textwidth]{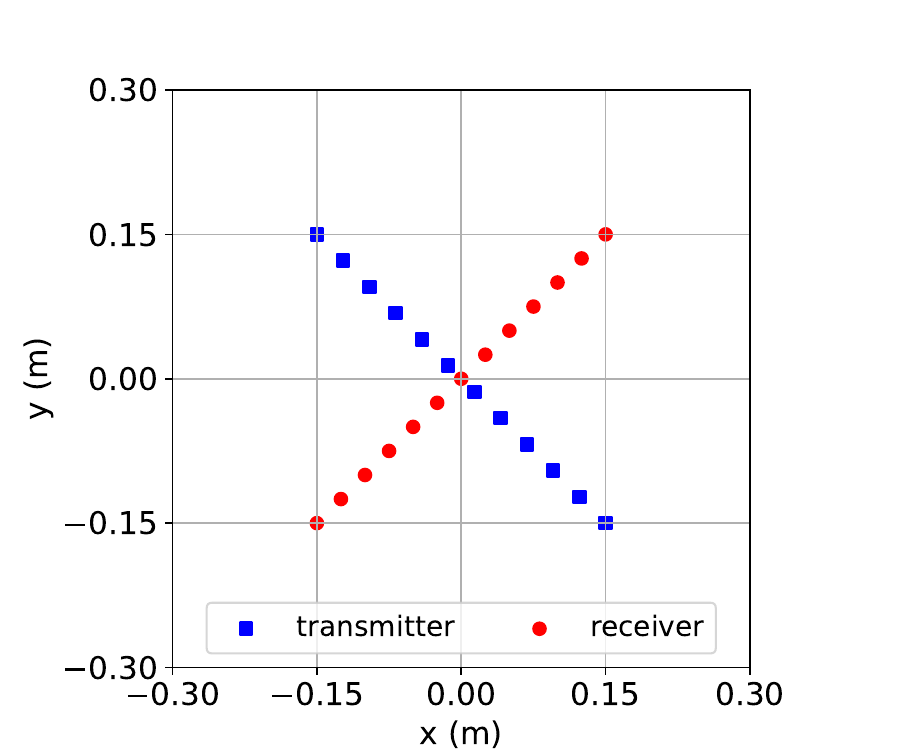}
    \caption{Mill's Cross Array.}
    \label{fig:CrossArray}
\end{figure}

As MIMO array topology, commonly used  Mill’s Cross array \cite{Zhuge2012RMA} is utilized. The used planar array has a width of 0.3 m, and contains 12 transmit and 13 receive antennas, which are uniformly spaced on the diagonals in a cross configuration as shown in Fig.~\ref{fig:CrossArray}. The frequency, $f$, is swept between 4 GHz and 16 GHz with uniform steps. 

For non-sparse measurement case with these aspects, the expected theoretical resolution~\cite{Zhuge2012RMA} is 2.5 cm in the cross-range directions, $x$ and $y$, and 1.25 cm in the down-range direction, $z$. With the goal of achieving these resolutions in the sparse case, we choose the image voxel size as 1.25 cm along $x$, $y$ directions, and 0.625 cm along $z$ direction (i.e. half of these resolutions). For the scene of interest, this results in an image cube of $25 \times25 \times 49$ voxels, which is same as the size of the synthetic scenes generated. Using these synthetic image cubes with the forward model in \eqref{eqn:forward-model-MVprod}, we simulate measurements at various signal-to-noise ratios ($\text{SNR}~=~10\log_{10}(\frac{\|\A\s\|^2_2}{M\cdot\sigma^2_{w}})$ ) and compression levels~($\text{CL}~=~1~-~\frac{M}{N}$) for our analysis.  

Before discussing the results, we provide the implementation details of the developed learning-based PnP approach, as well as the approaches used for comparison. For all regularization-based approaches, we enforce regularization on the reflectivity magnitudes and utilize the developed PnP approach in Algorithm~\ref{alg:ProposedMethod} with different denoising (proximal update) steps. In particular, as the proximal operator, $\mathbf{\Psi}_{\alpha \mathcal{R(\cdot)}}$, we utilize soft-thresholding in the case of $\ell_1$ regularization and 5 iterations of Chambolle algorithm \cite{chambolle2011first,Sidky2012} in the case of TV regularization. Although there are methods in the literature to decide on the value $\rho_2$ (or equivalently $\alpha$) adaptively, these methods introduce additional internal parameters to tune and can even negatively affect the convergence properties of the ADMM algorithm~\cite{boyd2011admm}. Here we choose the regularization parameter $\alpha$ in (27) by searching for its optimal value in the validation dataset between $10^{-5}$ and $10^{-1}$ in a coarse to fine fashion. We initialize each iterative algorithm with $\s^0~=~\frac{\A^H\y}{\max(|\A^H\y|)}$, and in each $\s$-update-step, the conjugate gradient algorithm is run for 5 iterations. TV and $\ell_1$-based approaches converge to a solution for a sufficiently large $\kappa$ in  \eqref{eqn:x-update}. Accordingly, we choose $\kappa=5\cdot10^4$ and run the iterations until the stopping criterion is satisfied, which is when the relative change $\frac{\||\s|^{l+1}-|\s|^{l}\|_2}{\||\s|^{l}\|_2}$ drops below $5\cdot10^{-4}$.
Because the convergence of learned PnP is an ongoing area of research and is not always guaranteed \cite{Ryu2019SpectralNormalization,Chan2017PnPADMM}, we limit the maximum number of iterations in the developed learning-based approach to 30. For the choice of $\kappa$, we search the optimal value using the validation dataset and set it as $\kappa=5\cdot 10^2$.

\begin{figure}
\centering
\begin{subfigure}[t]{0.45\linewidth}
\centering
Ground Truth\\
\includegraphics[width=1\linewidth]{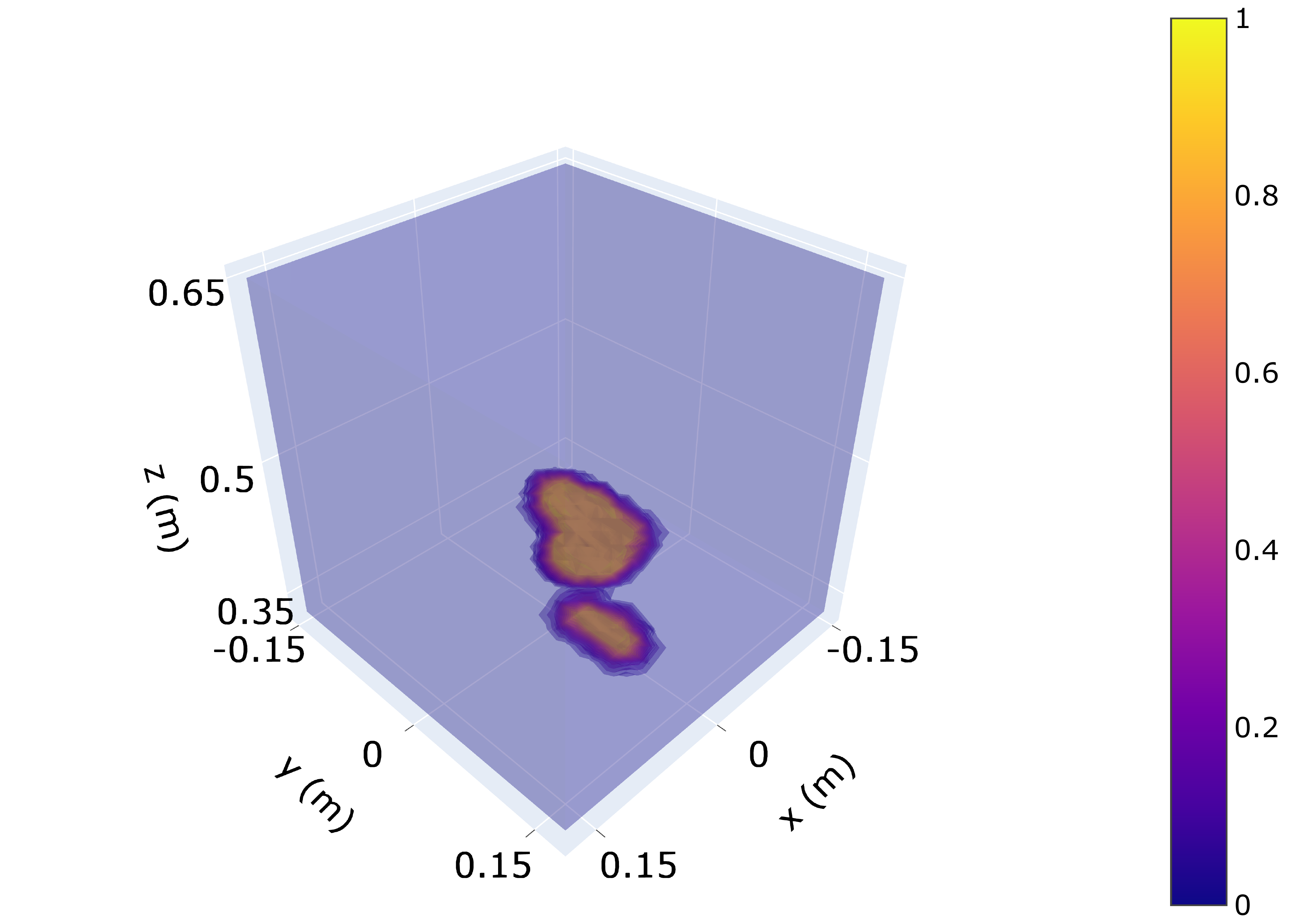}
\caption{}
\label{fig:gt}
\end{subfigure}

\newcommand{\tabfig}[1]{ \adjustbox{valign=m,vspace=1pt}{\includegraphics[width=1.5\linewidth]{#1}}}
\centering
\small
 \hspace*{-0.7cm}
 \begin{tabular}{m{1.4cm}m{1.4cm}m{1.4cm}m{1.4cm}m{1.4cm}}
\centering

\parbox{2.1cm}{\centering BP} & \parbox{2.1cm}{\centering KM} & \parbox{2.1cm}{\centering {$\ell_1$}} & \parbox{2.1cm}{\centering TV} & \parbox{2.1cm}{\centering Proposed} \\
\tabfig{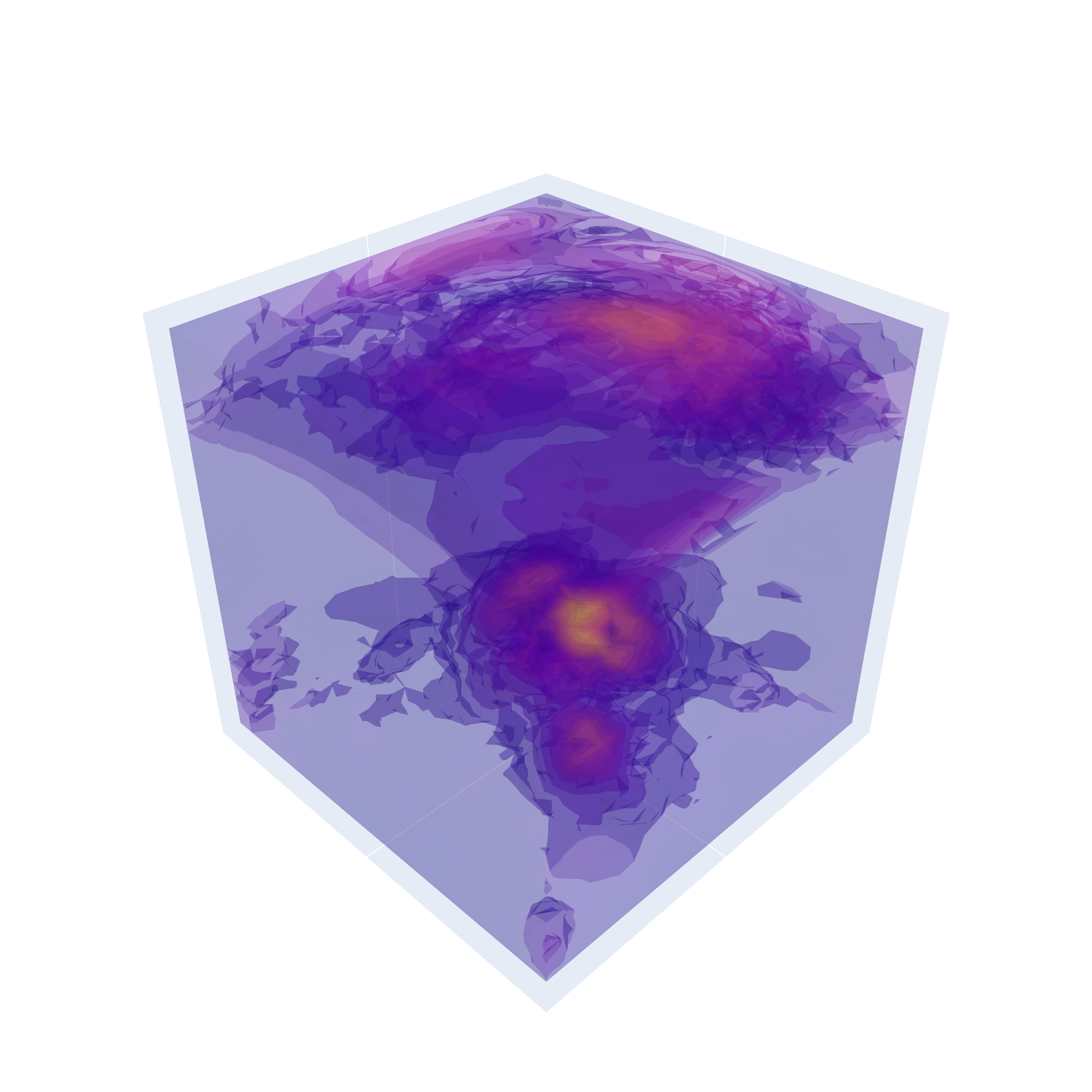} &\tabfig{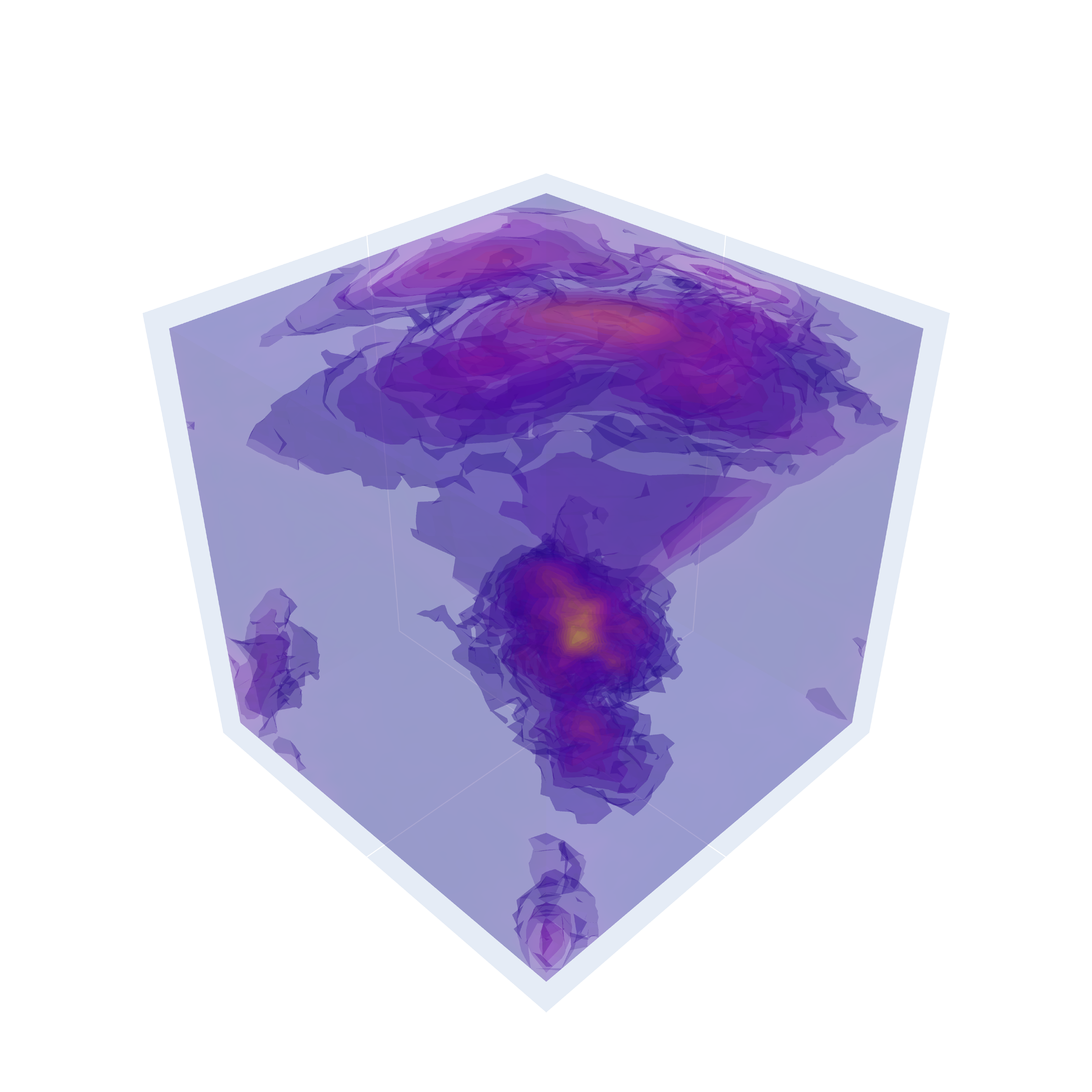} & \tabfig{figures/simulation-results/R2-S30-l1cg-csalsa-4.pdf} & \tabfig{figures/simulation-results/R2-S30-tvcg-csalsa-4.pdf} &\tabfig{figures/simulation-results/R2-S30-nncg-csalsa-4.pdf}\\
 \parbox{2.1cm}{\centering{(b)} 20.76 dB} & \parbox{2.1cm}{\centering{(c)} 22.08 dB} & \parbox{2.1cm}{\centering{(d)} 25.42 dB} & \parbox{2.1cm}{\centering{(e)} 26.13 dB} & \parbox{2.1cm}{\centering{(f)} 30.12 dB}  \\

\end{tabular}
    \caption{Sample reconstructions with $\frac{M}{N}~=~10\%$ data (i.e. 90$\%$ compression level) and 30 dB measurement SNR. (a) Ground truth, (b)-(f) Reconstructions obtained using different methods with their PSNR (dB) indicated underneath each figure. (Maximum projections along each dimension and 3D rotating views are available for all reconstructions at
https://github.com/METU-SPACE-Lab/PnP-Regularization-on-Magnitude as video.)}
\label{fig:initialTests}
\end{figure}

To comparatively evaluate the performance of the developed approach, we first consider the case with a medium SNR of 30 dB and a high compression level of 90$\%$. This corresponds to using 20 frequency steps between 4 and 16 GHz and is equivalent to reconstructing the reflectivity cube
with only 10$\%$ data. For a sample test image, the reconstructions obtained with different approaches are illustrated in Fig. \ref{fig:initialTests} using the same colormap. To quantitatively evaluate the performance, we also provide 3D peak signal-to-noise ratio (PSNR) between the normalized reconstructed magnitudes, $\frac{|\hat{\mathbf{s}}|}{\max|\hat{\mathbf{s}}|}$, and the ground truth magnitudes, $|\mathbf{s}|$, which is calculated as PSNR$~=~10\log_{10}\left( \frac{1}{\text{MSE}}\right)$ where MSE$~=\frac{1}{N} \| \; | \mathbf{s}| - \frac{|\hat{\mathbf{s}}|}{\max|\hat{\mathbf{s}}|} \|^2_2$ is the mean squared error. Although all algorithms reconstruct a complex-valued reflectivity distribution, the reconstructed phase is not used in this evaluation since it is random and does not contain any useful information. 
 As seen in Fig.~\ref{fig:initialTests}, the developed learning-based approach provides the best image quality with a reconstruction closely resembling the ground truth and achieving a PSNR of 30.12 dB. On the other hand, TV reconstruction suffers from over-smoothing, whereas $\ell_1$ based reconstruction contains speckle-like artifacts and an artifact cluster at the top. The visual quality of KM and BP reconstructions are even worse with many more reconstruction artifacts due to noisy and compressed data, where KM performs slightly better than BP.

To compare the reconstruction speed, average run-time of each method is computed over 100 test scenes as given in Table~\ref{tab:AverageRuntime}. As seen, the developed approach is capable of providing the best reconstruction quality with an average runtime of few seconds and is the fastest method after the direct inversion-based approaches (which largely fail). Moreover, TV and $\ell_1$ regularized solutions take much longer time to compute. 

\begin{table}
\caption{Average Run-Time on 100 Test Scenes at 30 dB SNR and with 10$\%$ data.} 
\centering
\begin{tabular}{c|ccccc}
\toprule
 & BP & KM & $\ell_1$ & TV & Proposed \\ \midrule
$\Delta t$ & 13.4 ms & 13.4 ms & 29.6 s & 21.2 s & 3.66 s\\\bottomrule
\end{tabular}
\label{tab:AverageRuntime}
\end{table}

\subsubsection{Compression Level Analysis}
We now analyze the effect of the compression level on the performance for the 30 dB SNR case. We consider compression levels of 97.5$\%$, 95$\%$, 92.5$\%$, 90$\%$, 85$\%$ and 80$\%$, which respectively correspond to using 5, 10, 15, 20, 30, and 40 frequency steps between 4 and 16 GHz, and are equivalent to reconstructing the reflectivity cube with 2.5$\%$, 5$\%$, 7.5$\%$, 10$\%$, 15$\%$ and 20$\%$ available data. Here the compression level of 97.5$\%$ is provided to show the breaking point of the proposed approach. For each case, the average PSNR is computed for the 100 test scenes reconstructed and is given in Table \ref{tab:CompressionAnalysis}. 

 As seen from the table, the developed learning-based approach significantly outperforms the other approaches for all compression levels other than 97.5$\%$ (i.e. 2.5$\%$ data). In particular, the average PSNR exceeds 30 dB when we perform a reconstruction with 10$\%$ or higher data. It is also interesting to observe that the performance of the developed method at the 95$\%$ compression level (i.e. 5$\%$ data) with 27.27 dB PSNR is even better than the performance of all compared methods at the lowest compression level (i.e. 20$\%$ data). At the 97.5$\%$ compression level with only 2.5$\%$ available data all methods fail to provide faithful reconstructions with PSNRs less than 23 dB, which suggests that the information provided by this amount of data is insufficient. As expected, all regularization-based approaches outperform the direct inversion methods (BP and KM), especially at highly compressive settings. Moreover, data-adaptive deep priors enable superior performance compared to hand-crafted analytical priors, TV, and $\ell_1$. From these analytical priors, $\ell_1$ starts to yield better performance than TV at the compression levels higher than 90$\%$ (i.e. with less than 10$\%$ data availability). From the direct inversion-based methods, KM consistently performs better than BP and approaches the performance of $\ell_1$ regularization at the increased data availability rates. Because of this, from this point forward, we will omit the BP from the visual comparisons and only present the results of KM. In general, the performance of each method starts to increase slowly with the increased data availability rates beyond 15$\%$. This suggests that the bottleneck on the measurement diversity becomes the sparse MIMO array topology when the number of frequency steps exceeds 30.

\begin{table}[]
\caption{Average PSNR on 100 Test Scenes for Different Amounts of Available Data at 30 dB Measurement SNR.}
\centering
\begin{tabular}{c|cccccc}
\toprule
$\frac{M}{N}$ & 2.5\% & 5\% & 7.5\% & 10\% & 15\% & 20\% \\ \midrule
Back-Projection &  16.41 & 19.75 & 21.71 & 23.49 & 24.56 & 24.60 \\ 
Kirchhoff Migration &  18.43 & 21.18 & 22.95 & 24.51 & 25.41 & 25.42 \\
$\ell_1$ Regularization&  22.76 & 24.08 & 24.90 & 25.70 & 25.85 & 25.85 \\ 
TV Regularization &  19.20 & 22.26 & 24.18 & 26.26 & 26.45 & 26.46 \\ 
Proposed Method &  21.53 & 27.27 & 29.82 & 30.40 & 30.65 & 30.75 \\ \bottomrule
\end{tabular}
\label{tab:CompressionAnalysis}
\end{table}

\begin{figure}[]
\newcommand{\tabfig}[1]{ \adjustbox{valign=m,vspace=1pt}{\includegraphics[width=1.5\linewidth]{#1}}}
\centering
\small
\hspace*{-0.65cm}\begin{tabular}{m{0.2cm}|m{1.4cm}m{1.4cm}m{1.4cm}m{1.4cm}}
\centering

$\frac{M}{N}$ & \parbox{2.1cm}{\centering KM} & \parbox{2.1cm}{\centering {$\ell_1$}} & \parbox{2.1cm}{\centering TV} & \parbox{2.1cm}{\centering Proposed} \\\midrule

\rotatebox{90}{20$\%$}& \tabfig{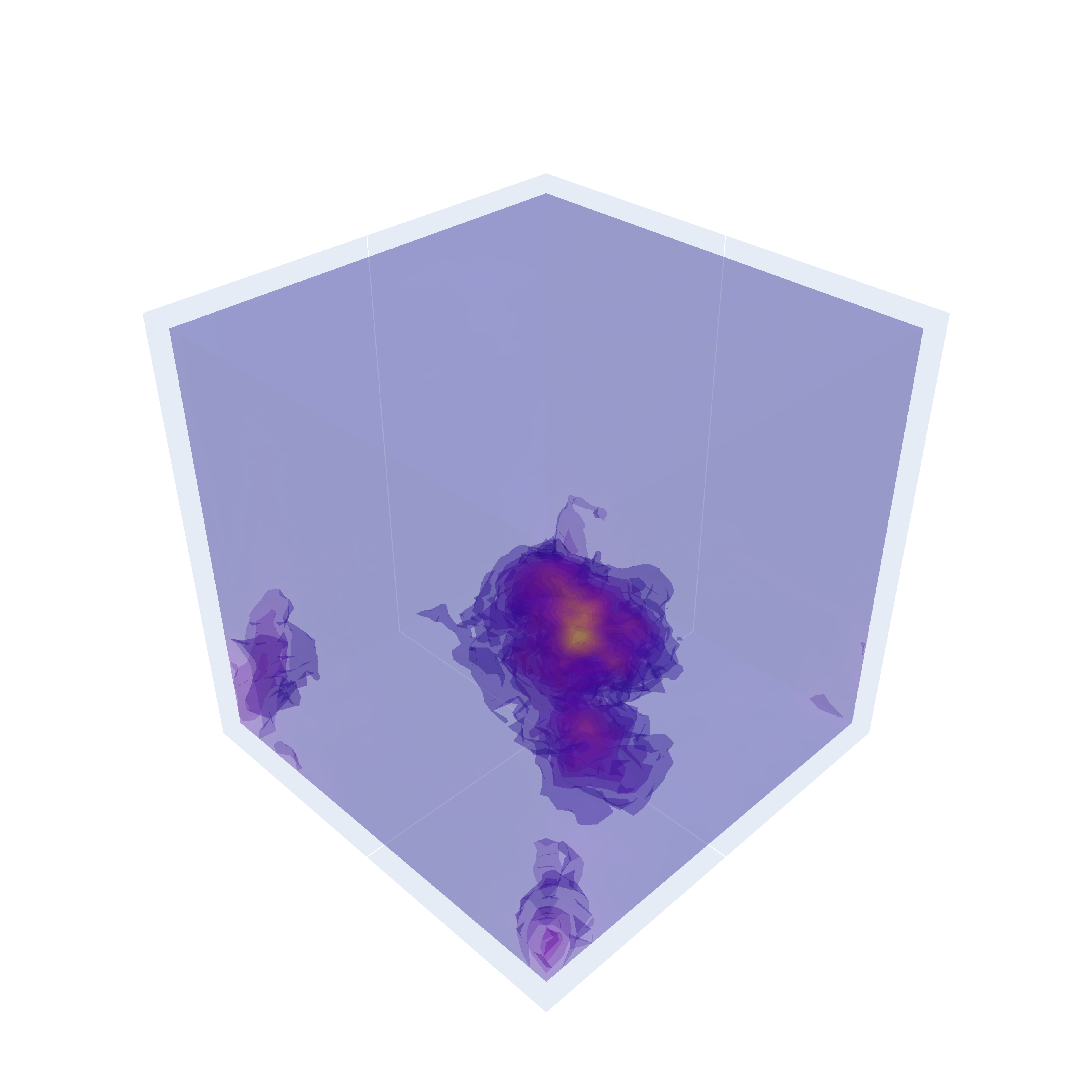} & \tabfig{figures/simulation-results/R4-S30-l1cg-csalsa-4.pdf} & \tabfig{figures/simulation-results/R4-S30-tvcg-csalsa-4.pdf} &\tabfig{figures/simulation-results/R4-S30-nncg-csalsa-4.pdf}\\
& \parbox{2.1cm}{\centering{(a)} 25.02 dB} & \parbox{2.1cm}{\centering{(b)} 25.42 dB} & \parbox{2.1cm}{\centering{(c)} 26.10 dB} & \parbox{2.1cm}{\centering{(d)} 31.07 dB} \\

\rotatebox{90}{10$\%$} &\tabfig{figures/simulation-results/R2-S30-kirchhoff-4.pdf} & \tabfig{figures/simulation-results/R2-S30-l1cg-csalsa-4.pdf} & \tabfig{figures/simulation-results/R2-S30-tvcg-csalsa-4.pdf} &\tabfig{figures/simulation-results/R2-S30-nncg-csalsa-4.pdf}\\
& \parbox{2.1cm}{\centering{(e)} 22.08 dB} & \parbox{2.1cm}{\centering{(f)} 25.42 dB} & \parbox{2.1cm}{\centering{(g)} 26.13 dB} & \parbox{2.1cm}{\centering{(h)} 30.12 dB} \\

\rotatebox{90}{5$\%$} & \tabfig{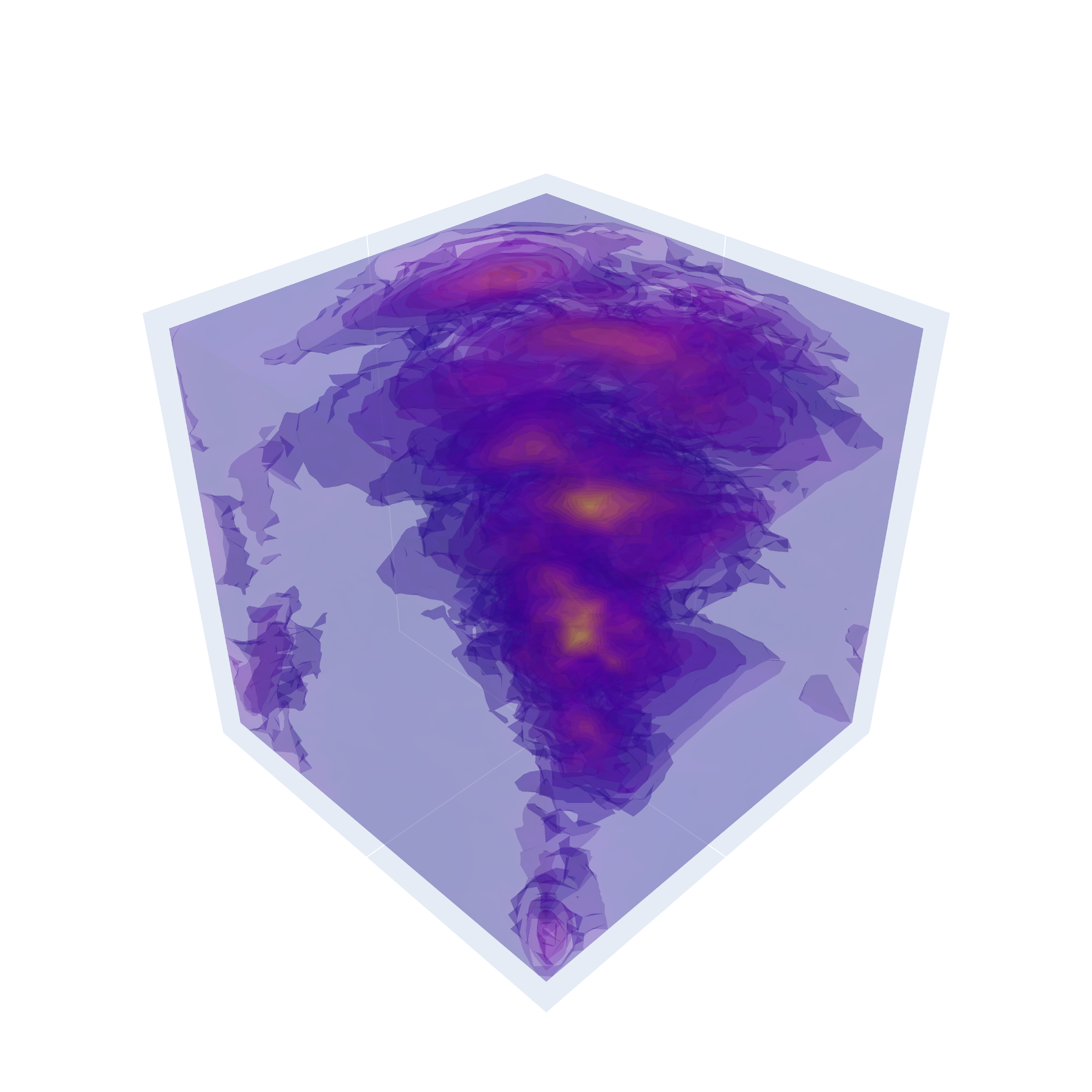} & \tabfig{figures/simulation-results/R1-S30-l1cg-csalsa-4.pdf} & \tabfig{figures/simulation-results/R1-S30-tvcg-csalsa-4.pdf} &\tabfig{figures/simulation-results/R1-S30-nncg-csalsa-4.pdf}\\
& \parbox{2.1cm}{\centering{(i)} 20.13 dB} & \parbox{2.1cm}{\centering{(j)} 24.71 dB} & \parbox{2.1cm}{\centering{(k)} 23.56 dB} & \parbox{2.1cm}{\centering{(l)} 29.54 dB} \\

\rotatebox{90}{2.5$\%$} & \tabfig{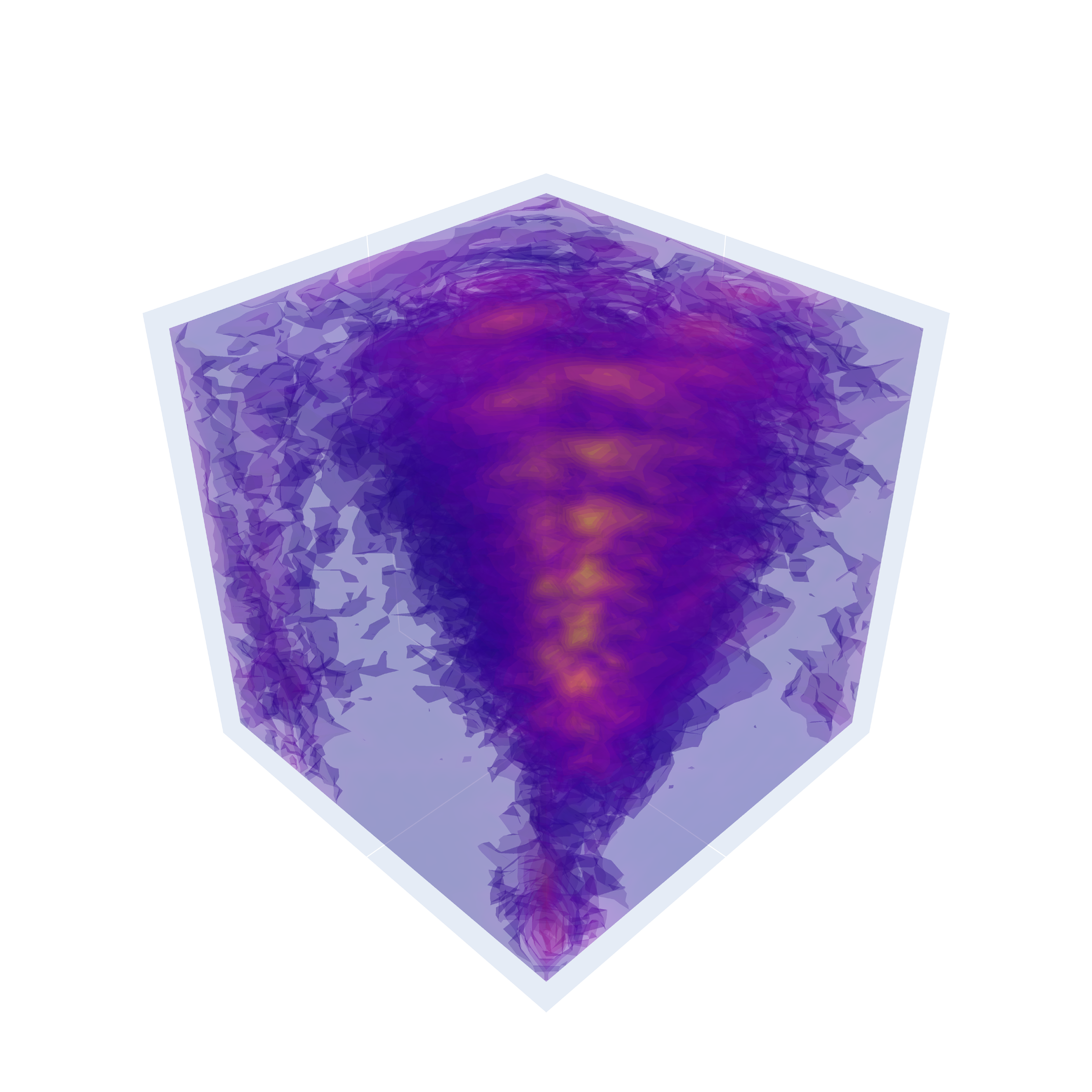} & \tabfig{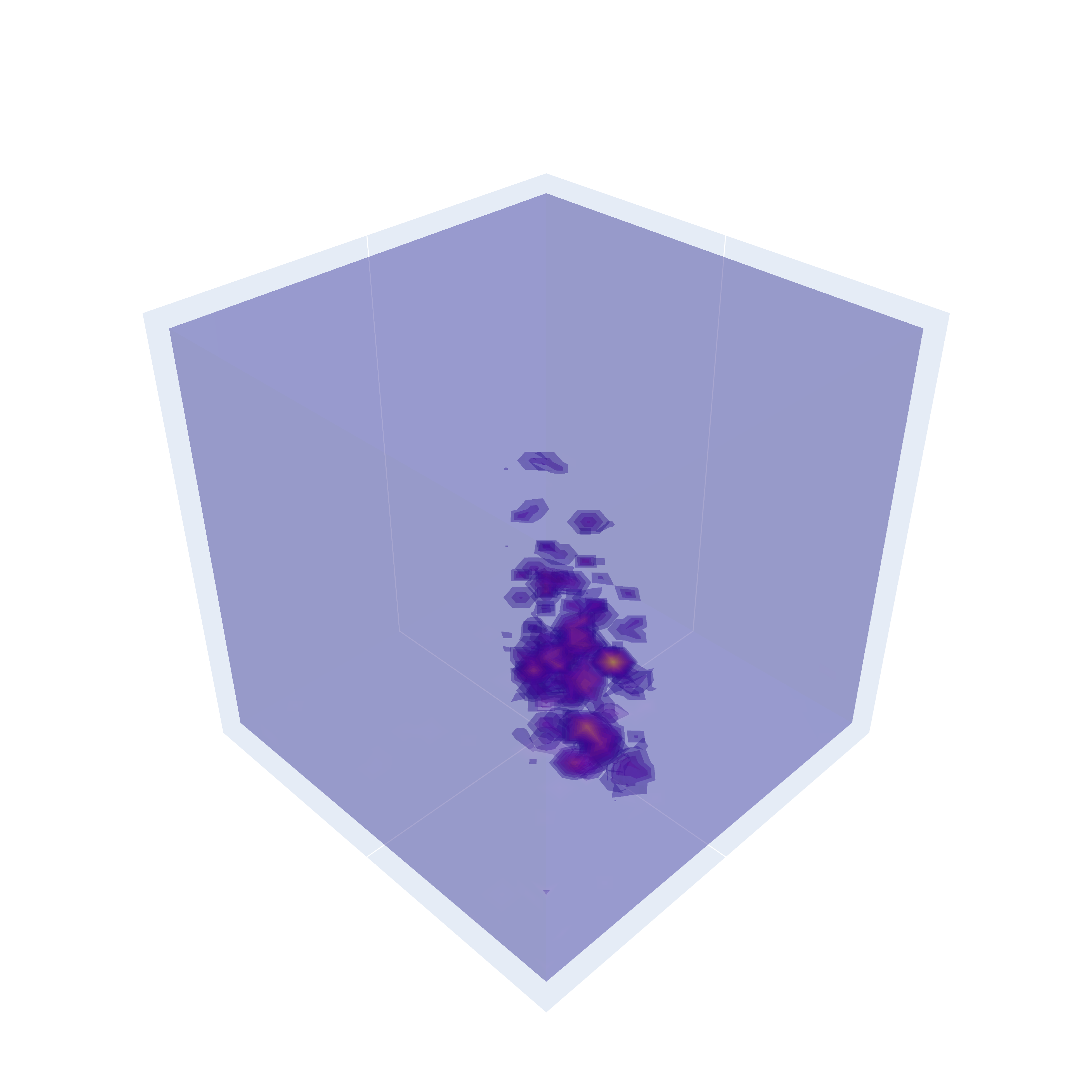} & \tabfig{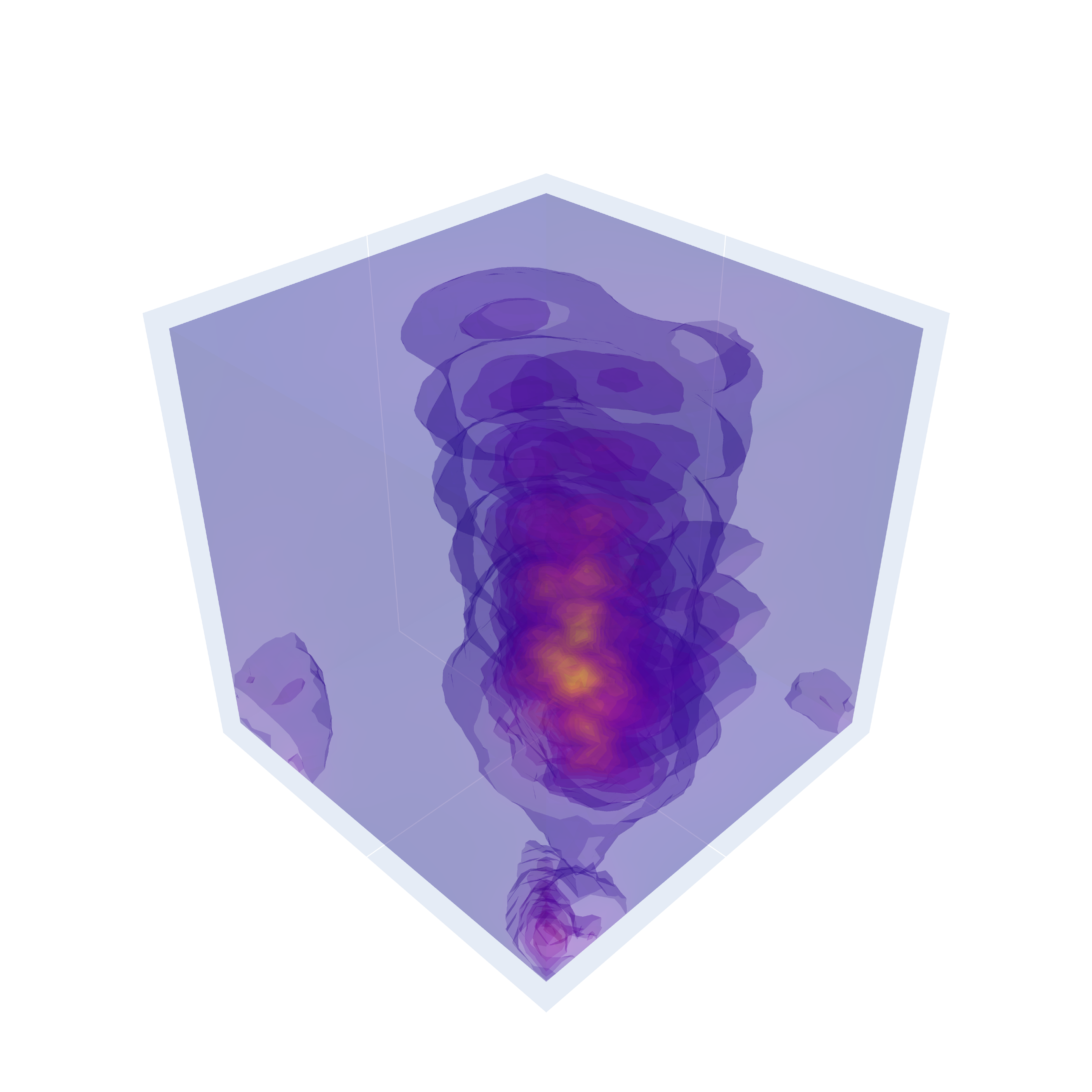} &\tabfig{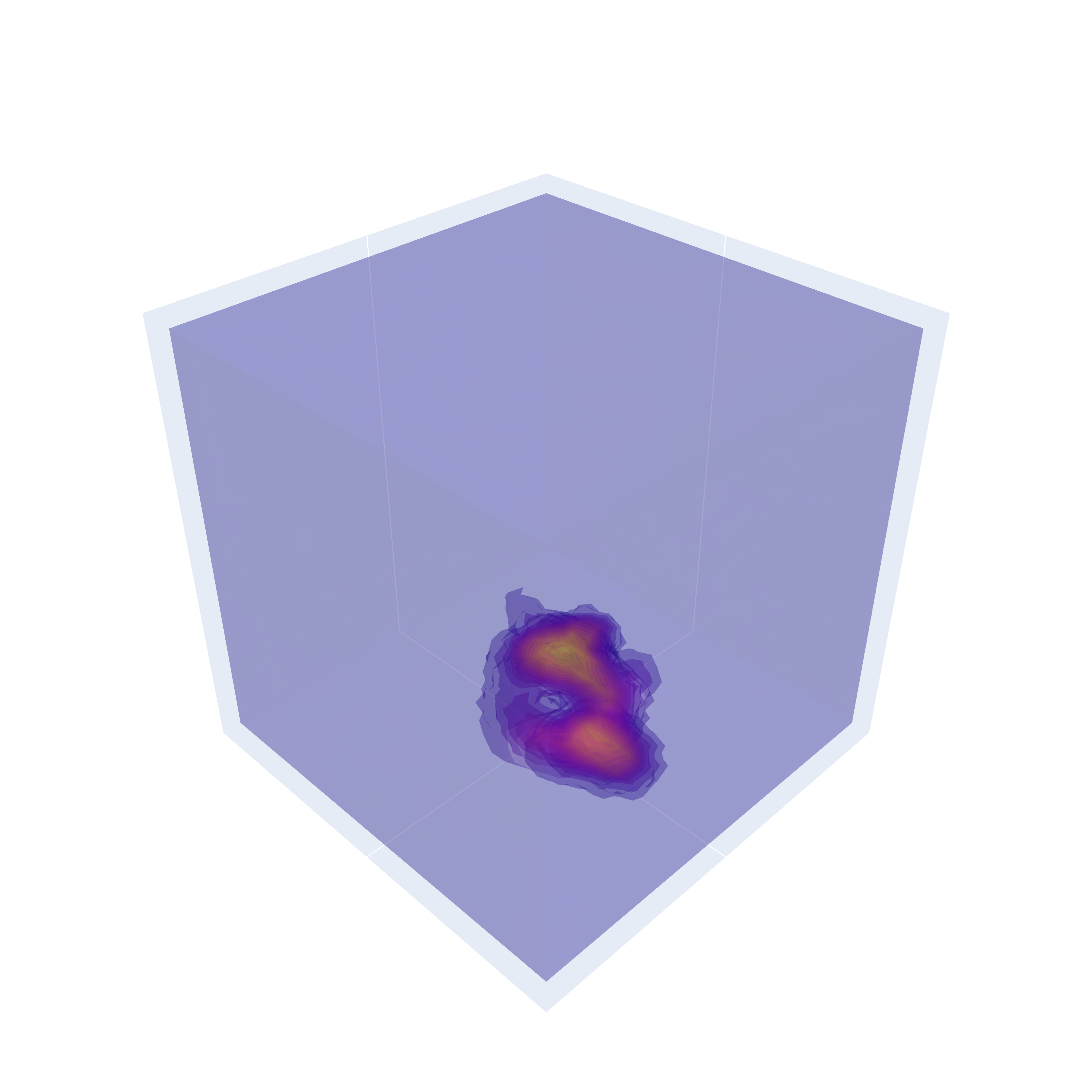}\\
& \parbox{2.1cm}{\centering{(m)} 17.09 dB} & \parbox{2.1cm}{\centering{(n)} 22.54 dB} & \parbox{2.1cm}{\centering{(o)} 19.98 dB} & \parbox{2.1cm}{\centering{(p)} 21.77 dB} \\

\end{tabular}
    \caption{Sample reconstructions obtained for different amounts of available data, $\frac{M}{N}~=$$~2.5\%,$$5\%,10\%,20\%$, at 30 dB measurement SNR. PSNR (dB) of each reconstruction is indicated underneath.}
    \label{fig:CompessionTest}
    \end{figure}

For visual comparison, sample reconstructions obtained with 2.5$\%$, 5$\%$, 10$\%$, and 20$\%$ data are also given in Fig.~\ref{fig:CompessionTest}. As seen, for all approaches, the reconstruction quality improves with the increasing amount of data. Moreover, we can observe that KM is the most severely affected method by the amount of available data, and at high compression levels its reconstruction suffers from large grating lobes. At compression levels corresponding to 5$\%$ and 2.5$\%$ data, TV reconstruction also contains large artifacts in addition to the over-smoothing effect. On the other hand, although the $\ell_1$-based method suffers from speckle-like artifacts, its performance does not change much up until the highest compression level (corresponding to 2.5$\%$ data). For the highest compression level, we see that $\ell_1$ reconstruction is point-like and not an extended target. On the other hand, although the PSNR of the developed method is less, it outputs an extended target that resembles the shape of the ground truth. After this breaking point for the compression level, the proposed learning-based PnP method yields almost artifact-free reconstructions for all other compression levels.

\subsubsection{Noise Level Analysis}
We now fix the available data to 10$\%$ and analyze the effect of SNR on the quality of reconstructions. For this, we gradually drop the SNR from 30 dB to 0 dB with steps of 10 dB. The average PSNR of each method is given in Table \ref{tab:SNRAnalysis} at different SNRs. As seen, the developed learning-based approach outperforms the other methods also for all noise levels. In particular, the performance of the developed method even at the lowest SNR case (i.e. 0 dB) with 28.31 dB PSNR is better than the performance of all compared methods at the highest SNR case (i.e. 30 dB). Similar to the results in the compression level analysis, all regularization-based approaches outperform the direct inversion methods, and in the most ill-posed case with 0 dB SNR, $\ell_1$ prior yields better reconstruction than TV.

\begin{table}[]
\centering
\caption{Average PSNR on 100 Test Scenes for Different Measurement SNRs using 10$\%$ Data.} 
\begin{tabular}{c|cccc}
\toprule
SNR              &0 dB          & 10 dB         & 20 dB          & 30 dB         \\\midrule
Back-Projection      &22.20       & 23.35       & 23.48       & 23.49  \\
Kirchhoff Migration  & 22.37      & 24.26       & 24.49       & 24.51   \\
$\ell_1$ Regularization&25.40  & 25.69 & 25.70 & 25.70 \\
TV Regularization&23.87 & 26.02 & 26.26 & 26.26  \\
Proposed Method & \textbf{28.31}       & \textbf{29.28}       & \textbf{30.12 }      & \textbf{30.40 }     \\\bottomrule
\end{tabular}
\label{tab:SNRAnalysis}
\end{table}

\begin{figure}[]
\newcommand{\tabfig}[1]{ \adjustbox{valign=m,vspace=1pt}{\includegraphics[width=1.5\linewidth]{#1}}}
\centering
\small
 \hspace*{-0.70cm}\begin{tabular}{m{1.4cm}m{1.4cm}m{1.4cm}m{1.4cm}}
\centering

 \parbox{2.1cm}{\centering KM} & \parbox{2.1cm}{\centering {$\ell_1$}} & \parbox{2.1cm}{\centering TV} & \parbox{2.1cm}{\centering Proposed} \\
\tabfig{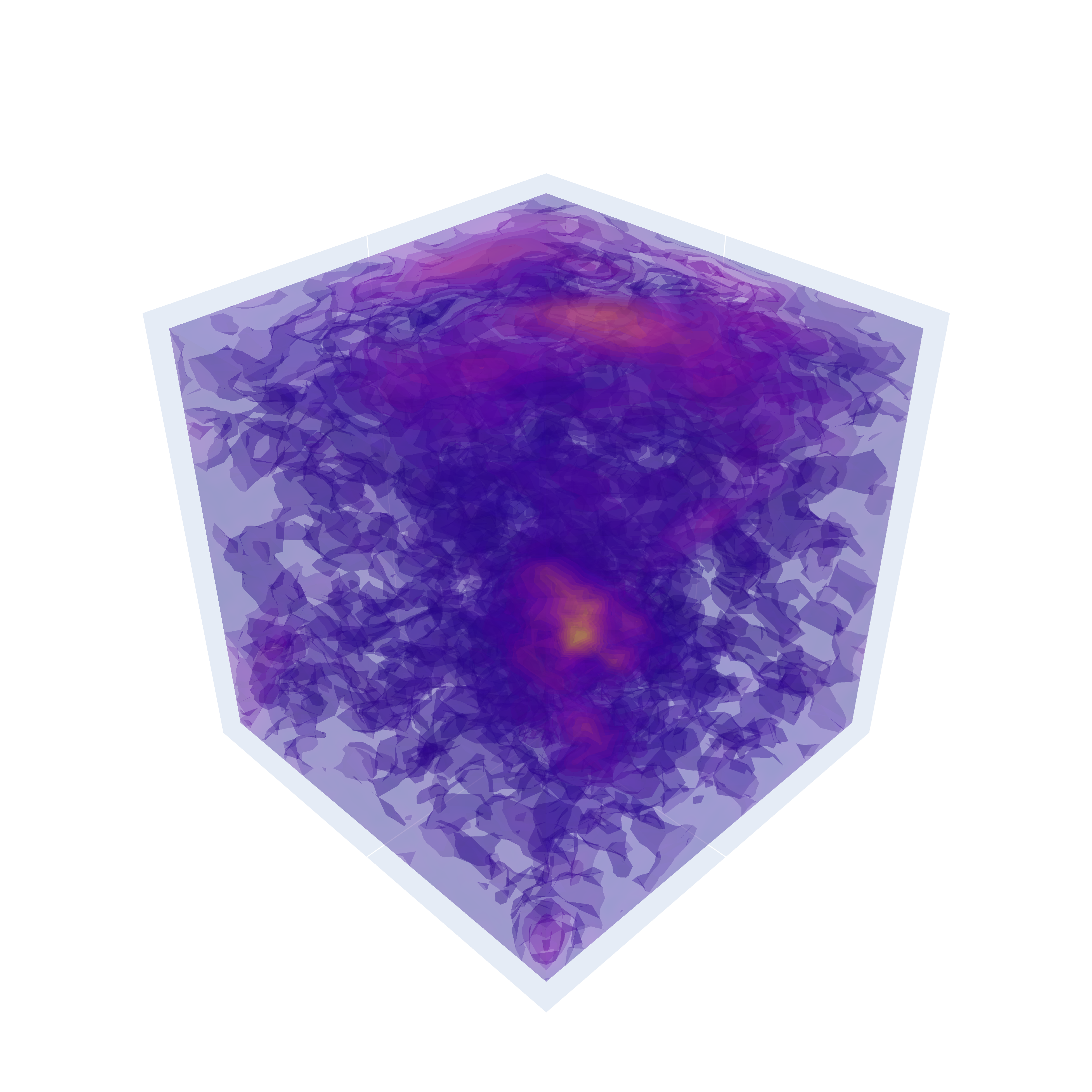} & \tabfig{figures/simulation-results/R2-S0-l1cg-csalsa-4.pdf} & \tabfig{figures/simulation-results/R2-S0-tvcg-csalsa-4.pdf} &\tabfig{figures/simulation-results/R2-S0-nncg-csalsa-4.pdf}\\
 \parbox{2.1cm}{\centering{(a)} 20.58 dB} & \parbox{2.1cm}{\centering{(b)} 24.89 dB} & \parbox{2.1cm}{\centering {(c)} 22.93 dB} & \parbox{2.1cm}{\centering{(d)} 28.88 dB}  \\

\end{tabular}
    \caption{Sample reconstructions with $\frac{M}{N}~=~10\%$ data at 0 dB measurement SNR. PSNR values are indicated underneath the figures. }
  \label{fig:SNRTest} 
\end{figure}

In Fig. \ref{fig:SNRTest} sample reconstructions for 0 dB SNR case are given. Compared to the reconstructions given in Fig. \ref{fig:initialTests} for 30 dB SNR case, KM result is severely degraded at this low SNR due to high noise amplification. On the other hand, $\ell_1$ and TV-based reconstructions still show some fidelity to the original image, but with more artifacts. More importantly, even for this highly noisy and compressive observation setting, the proposed learning-based PnP method is capable of providing a clean reconstruction that maintains high fidelity to the ground truth. 

\subsection{Performance Analysis with Experimental Data}

\begin{figure}[]
    \centering
    \includegraphics[width=0.30\textwidth]{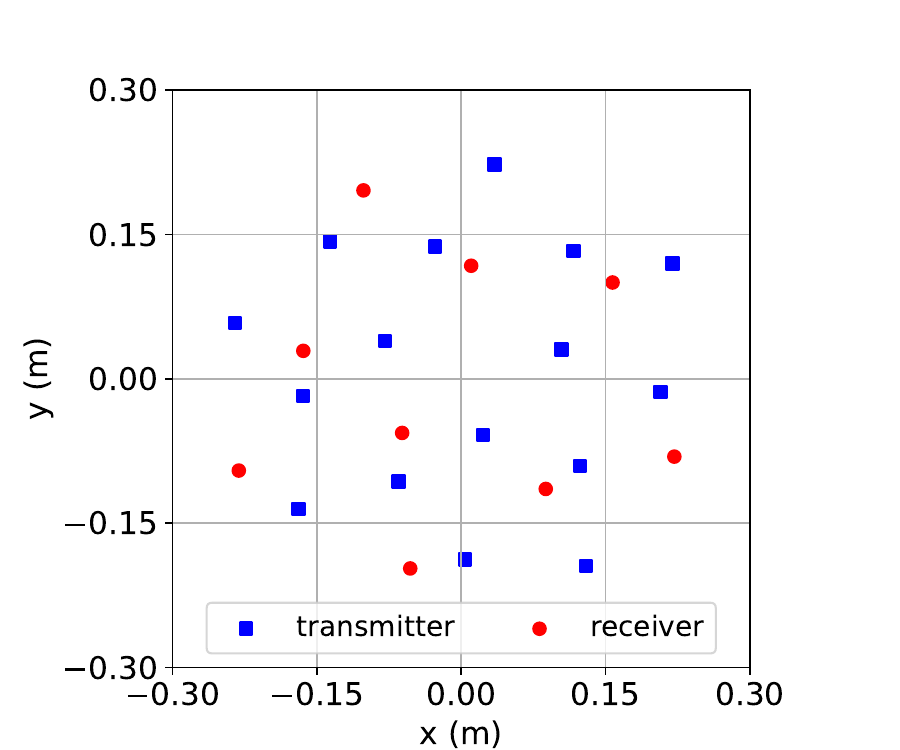}
    \caption{Spiral MIMO Array.}
    \label{fig:YarovoyArray}
\end{figure}

We now demonstrate the performance of the developed approach on real-world scenes using experimental measurements available online~\cite{Wang2020EMData,Wang2020Short-Range}. These experimental measurements were acquired for a scene that contains a toy revolver approximately 50 cm away from a sparse MIMO array~\cite{Wang2020Short-Range}. The used MIMO array has 16 transmit and 9 receive Vivaldi antennas that are distributed in a spiral configuration on the antenna plane as shown in Fig.~\ref{fig:YarovoyArray}. The experimental measurements were recorded at 251 uniformly sampled frequencies from 1 to 26 GHz. We aim to infer the reflectivity distribution within a 30 cm $\times$ 30 cm $\times$ 30 cm image cube that contains the revolver. Similar to \cite{Wang2020Short-Range}, we choose the sampling interval as 0.5 cm along all three dimensions. This results in an unknown image cube of $61 \times 61 \times 61$ voxels.

Since our focus is on compressive imaging, we consider sparse frequency measurements from the band of 4--16~GHz (similar to the simulated setting). In particular, from the available data, we use 7 and 11 uniformly sampled frequencies between 4 and 16~GHz, which respectively correspond to compression ratios of 99.56$\%$ and 99.31$\%$. These are equivalent to reconstructing the reflectivity cube with only 0.44$\%$ and 0.69$\%$ data, yielding to extremely compressive settings. 

To reconstruct this real scene using the developed approach with deep prior as well as with TV and $\ell_1$ priors, we use the same $\kappa$ parameters determined in the previous simulated setting. For the choice of the regularization parameter $\alpha$, we again perform a search for the optimal value to obtain the best reconstruction quality.  Moreover, the parameter $\epsilon$ in \eqref{eqn:proj} is empirically set to $\frac{1}{\sqrt{10}}\|y\|_2$, which approximately corresponds to measurement at 10 dB SNR. Additionally, since the maximum value of the reflectivity magnitudes in the real scene can be different from the synthetic scenes used in training, the reflectivity magnitude at each iteration is scaled with its maximum value prior to entering to the denoiser (in order to fall into the range $[0,1]$). Then the denoised magnitude at the output of the denoiser is scaled back. 

\begin{figure*}[t]
\newcolumntype{P}[1]{>{\centering\arraybackslash}p{#1}}
\newcolumntype{M}[1]{>{\centering\arraybackslash}m{#1}}
\newcommand{\tabfig}[1]{ \adjustbox{valign=m,vspace=3pt }{\includegraphics[width=1\linewidth]{#1}}}

\centering
\small
\begin{subfigure}[]{0.167\textwidth}
 \begin{tabular}{M{\linewidth}}
    \centering
     \vspace{6pt}Toy Revolver\vspace{5pt}\\
   \adjustbox{valign=m,vspace=0pt}{\includegraphics[width=1\linewidth]{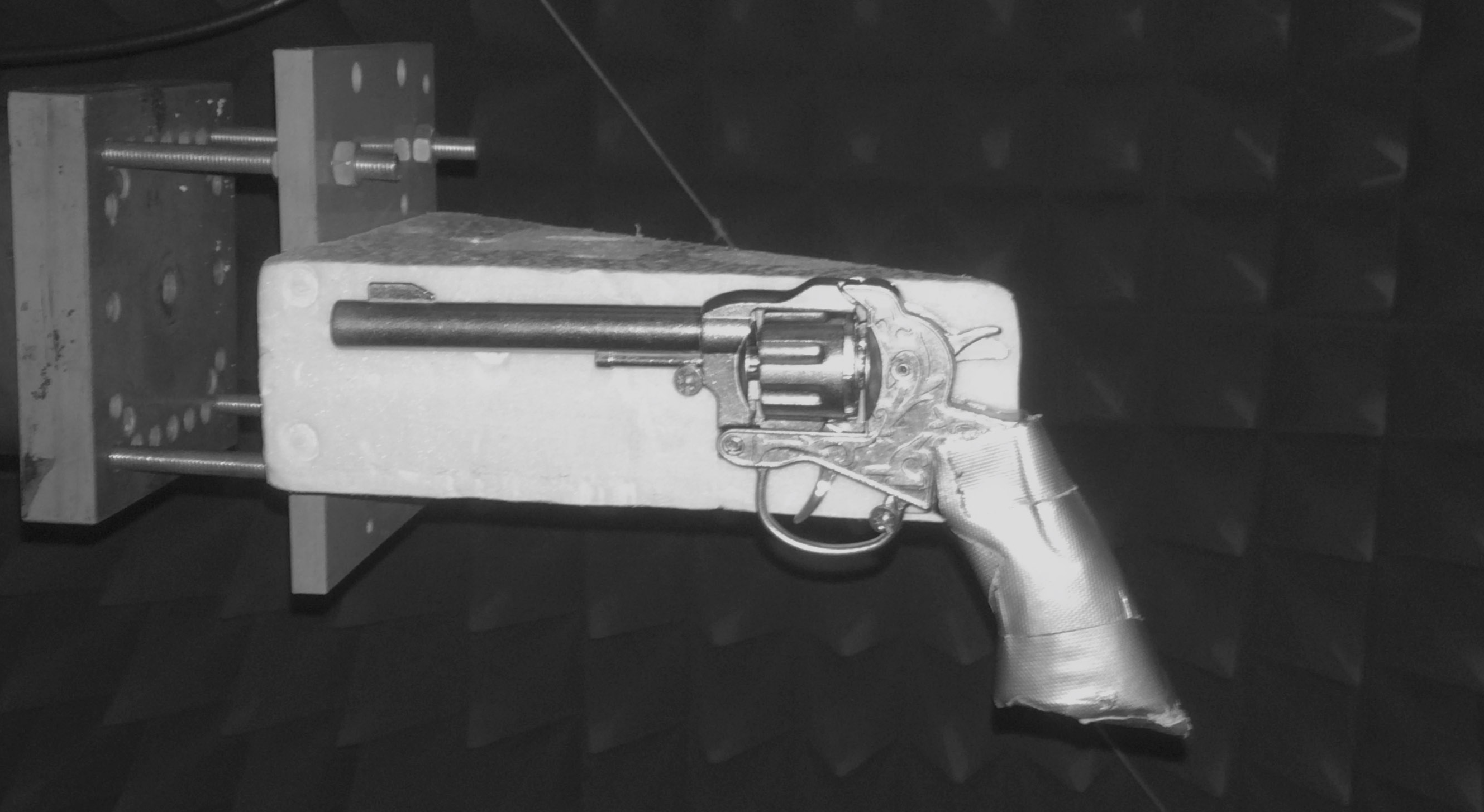}} \caption{\cite{Wang2020EMData,Wang2020Short-Range}} 
   \vspace{8pt} Full-Data Kirchhoff~Migration  \\
   \tabfig{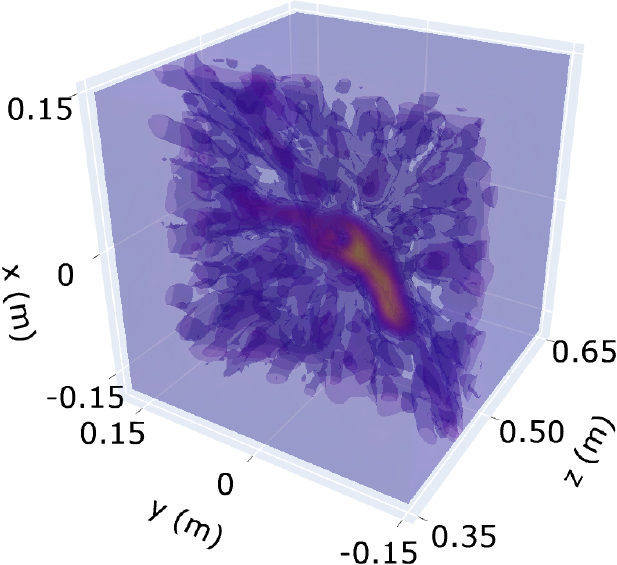}
   \caption{}
   \label{fig:FullDataKM}
\end{tabular}
\end{subfigure}
\hfill
\begin{subfigure}[]{0.8\textwidth}
    \begin{tabular}{|M{0.02\textwidth}|M{0.2\textwidth}M{0.2\textwidth}M{0.2\textwidth}M{0.2\textwidth}}
    \centering
    $\frac{M}{N}$ & Kirchhoff Migration &  TV Regularization & $\ell_1$ Regularization & Proposed Method \\\midrule

     \rotatebox{90}{0.44$\%$} & \tabfig{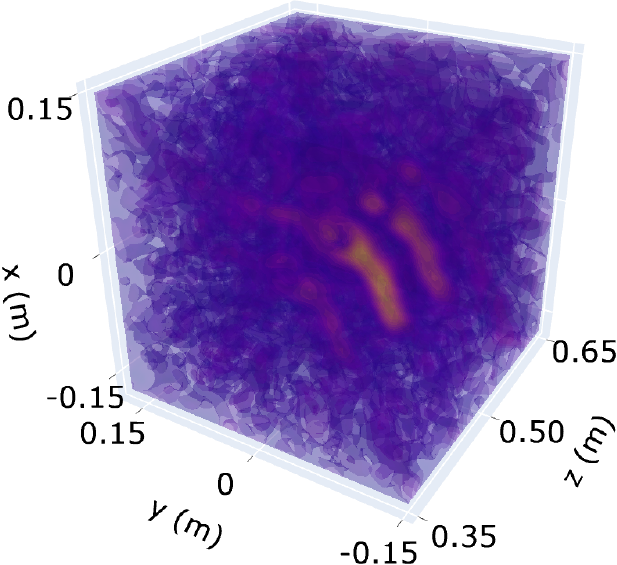} & \tabfig{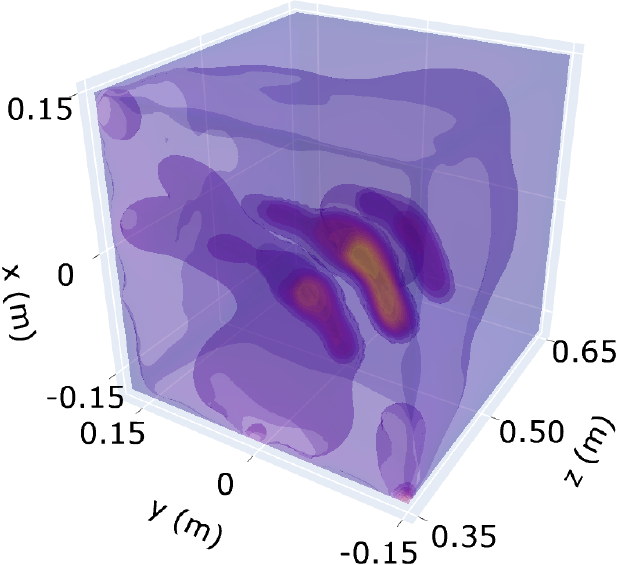} & 
     \tabfig{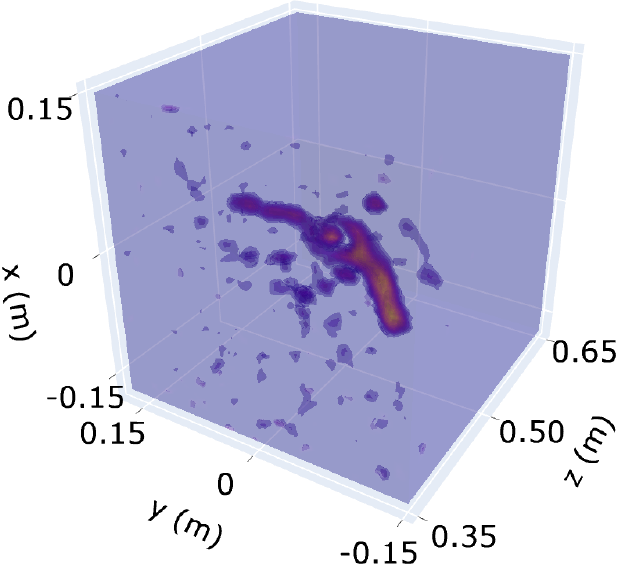} & \tabfig{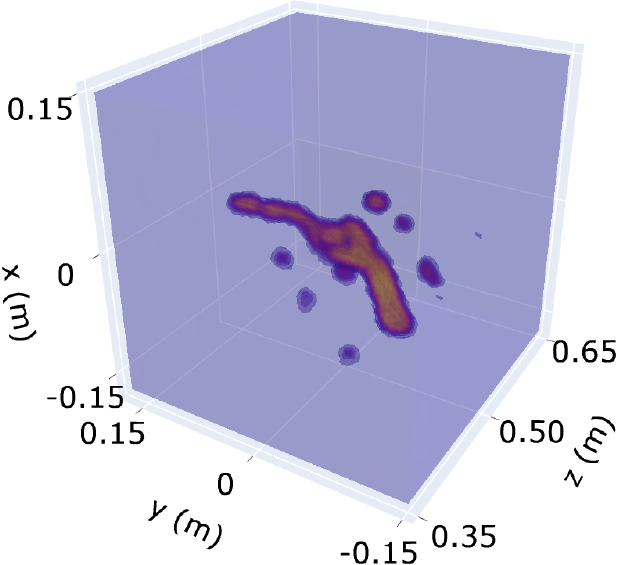} \\
     \rotatebox{90}{0.69$\%$} & \tabfig{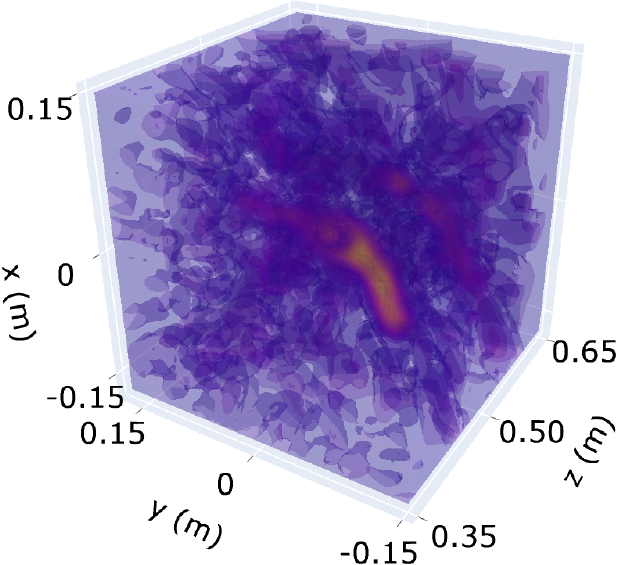} & \tabfig{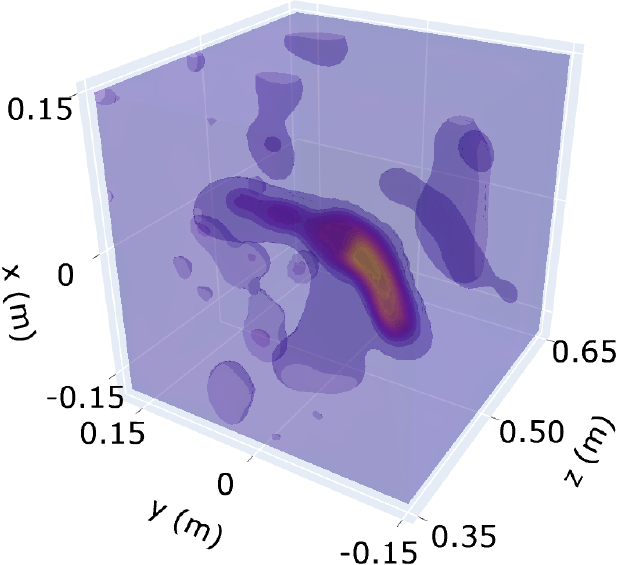} & 
     \tabfig{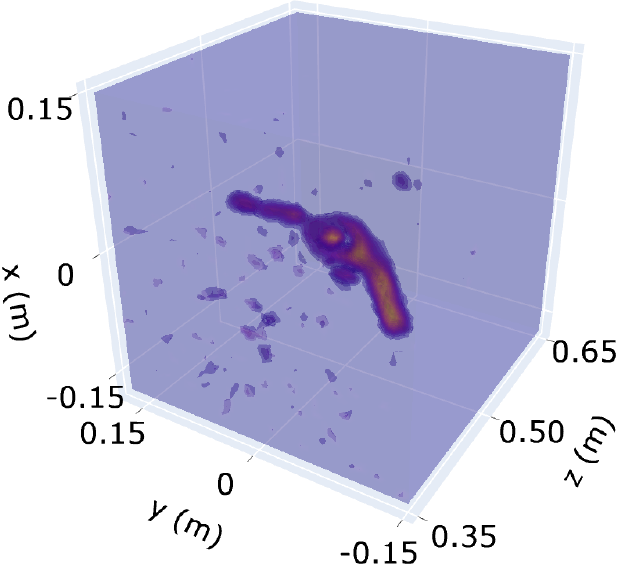} & \tabfig{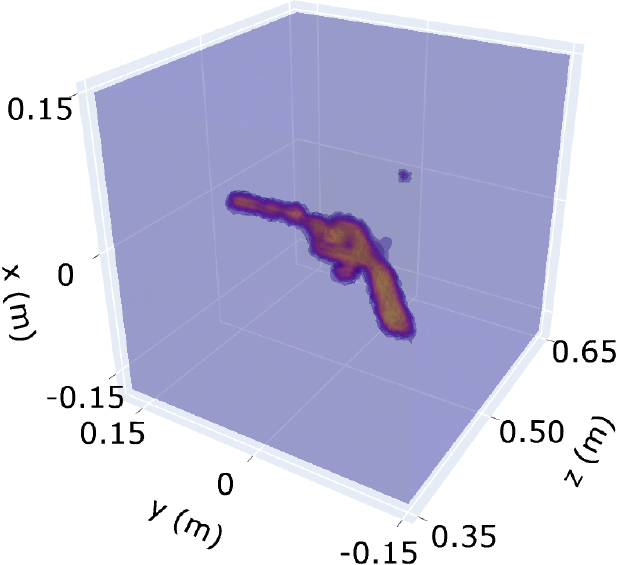} \\
\end{tabular}
\caption{}
\label{fig:YarovoyResults-compressive}
\end{subfigure}

\caption{Imaged revolver and its reconstructions with experimental data; (a) photograph of the toy revolver, (b)~\textit{full-data} ($\frac{M}{N}=$361.46$\%$) KM reconstruction, (c) reconstructions obtained with different methods at two compressive settings using 7 ($\frac{M}{N}=$0.44$\%$) and 11 ($\frac{M}{N}=$0.69$\%$) frequency steps. }
\label{fig:YarovoyResults}
\end{figure*}

A photograph of the imaged toy revolver and the reconstructions obtained for two different compressive settings with 0.44$\%$ and 0.69$\%$ data 
are shown in Fig.~\ref{fig:YarovoyResults}. Note that the photograph provides a visual reference for comparisons, but it does not represent the ground truth reflectivity magnitudes. As additional reference for comparisons, we also obtain the KM reconstruction of the scene using the full frequency data available (i.e. 251 frequency steps in the band 1-26 GHz), which corresponds to a highly over-determined setting with $\frac{M}{N}=361.44\%$ data availability. This \textit{full-data} KM reconstruction is given in Fig.~\ref{fig:FullDataKM} to reveal the general shape of the scene reflectivity. But despite using all of the available data, it still contains widespread artifacts, especially over the cross-range dimensions. This is the expected behavior of direct inversion methods with sparse arrays due to the resulting aliasing~\cite{Wang2020Short-Range}.

When we compare the reconstructions in Fig.~\ref{fig:YarovoyResults-compressive} for the highly compressive settings considered, it is seen that the developed approach with deep prior provides the best results with the least amount of artifacts. In particular, KM reconstructions suffer from significant grating lobes and aliasing on range direction (which appears in the form of replication) resulting due to the sparsely sampled frequencies. Although not as prominent, similar replication artifacts on range direction are also present in the results of hand-crafted regularization approaches. Most notably, TV reconstructions fail to resolve aliasing and contain replicated silhouettes of the revolver. While TV reconstructions perform visually better than KM, they perform poorly compared to $\ell_1$ regularization at these highly compressive settings (as similar with the observations in the earlier analysis). In $\ell_1$ regularized reconstructions, there are less artifacts along the cross-range directions compared to TV, but the revolver appears as eroded, and there are distributed speckle artifacts, which are more common along the range direction (aligned with the locations of the aliasing artifacts in KM- and TV-based solutions). 

On the other hand, the proposed PnP approach with deep prior is capable of providing a near-perfect reconstruction with only 0.69$\%$ data. Few aliasing artifacts occur over the range direction at the higher compressed setting with 0.44$\%$ data. Nevertheless, in both cases, the edges of the object are sharply reconstructed, and the frame, cylinder, trigger guard, and muzzle of the revolver are all clearly visible. Hence the proposed approach is much less prone to sparse sampling and aliasing, thanks to the power of learned deep priors. Note that this is in spite of the fact that the spatial resolution of the test object is higher compared to the training dataset. Higher resolution reconstructions can also be successfully obtained as illustrated in the provided supplementary document.

The proposed method not only provides the highest reconstruction quality but also takes only 6 seconds (for the case with $0.44\%$ data). Hence it is again the second fastest method after KM which performs poorly. On the other hand, TV and $\ell_1$ regularized solutions suffer from significantly longer computation time, which are approximately 150 seconds. 

Overall these real scene experiments demonstrate that the utilization of deep priors in a plug-and-play algorithm enables state-of-the-art reconstruction quality even at highly compressive experimental settings, while also yielding significantly reduced run-time compared to hand-crafted analytical priors. Note that the learned prior is also capable of representing unseen real-world objects, although the training has been performed with synthetic and randomly generated much simpler extended targets. Moreover, even though this experimental observation setting (including antenna array type, number of measurements taken, etc.) differs from the previously analyzed simulated setting, our learning-based method can be directly used without re-training since it is based on PnP framework (and not unrolling). Hence the proposed learning-based PnP method is highly adaptable to experimental data and different observation settings. 

\section{Conclusion}
\label{Section:Conclusion}

We have developed a novel and efficient plug-and-play approach that enables the reconstruction of 3D complex-valued images involving random phase by exploiting both analytic and deep priors. Our approach provides a unified general framework to effectively handle arbitrary regularization on the magnitude of a complex-valued unknown and is applicable to various complex-valued image formation problems including SAR and MIMO radar imaging with far- or near-field settings. Our development is based on a general closed-form expression provided for the solution of a complex-valued denoising problem with regularization on the magnitude. By utilizing this expression in an ADMM framework, a computationally efficient PnP reconstruction method that consists of simple update steps is obtained.

In this paper, we applied the developed PnP method to near-field compressive MIMO imaging for reconstruction of the 3D complex-valued scene reflectivities with random phase nature. Within our PnP framework, we utilized a 3D deep denoiser to take advantage of data-adaptive deep priors. To the best of our knowledge, our approach is the first deep prior-based PnP approach demonstrated for near-field radar imaging.

The effectiveness of our approach is illustrated under various compressive and noisy observation scenarios in microwave imaging using both simulated and experimental data. The results show that the developed PnP approach with learned deep prior achieves the state-of-the-art reconstruction quality at highly compressive settings with a generalizability capability for unseen real-world objects and high adaptability to experimental data. The approach also has the advantage of reduced run-time and applicability to different observation settings without re-training due to its PnP nature. Compared to approaches with analytical priors, it is also more robust to sparse data and noise. We observe both with simulated and experimental data that frequency steps as few as 10 provide sufficient measurement diversity for reconstruction of scenes with average complexity. This is an important observation since earlier works generally use hundreds of frequency steps for similar tasks. As expected the bandwidth is more critical than the number of frequency samples taken within this band. 

Lastly we note that although the developed PnP method is quite fast with a runtime on the order of seconds, further acceleration and reduction in memory use can be achieved by more efficiently computing the forward and adjoint operators, using methods like fast multipole method (FMM)~\cite{miran2021sparse}. Moreover, exploring the performance of the developed method with different 3D denoiser architectures, and joint optimization of the denoiser and MIMO array configuration may improve the reconstruction quality, which are topics for future study. Enriching our training dataset can also help to improve the performance. Likewise, utilizing a training dataset synthesized for a specific imaging task, such as a dataset consisting of 3D models of concealed weapons, can allow the deep architecture to better learn the task-oriented prior information and can improve the performance. 

\section*{Acknowledgments}
The authors thank professors Sencer Koc and Lale Alatan at METU for many fruitful discussions about radar imaging. 
This work is supported by the Scientific and Technological Research Council of Turkey (TUBITAK) under grant 120E505.
\bibliography{bibliography}
\bibliographystyle{IEEEtran}

\bigskip
\bigskip
\bigskip
\bigskip
\bigskip
\begin{minipage}[H]{2.5cm}
\vspace{-.2cm}
{\includegraphics[width=1in,height=1.25in,clip,keepaspectratio]{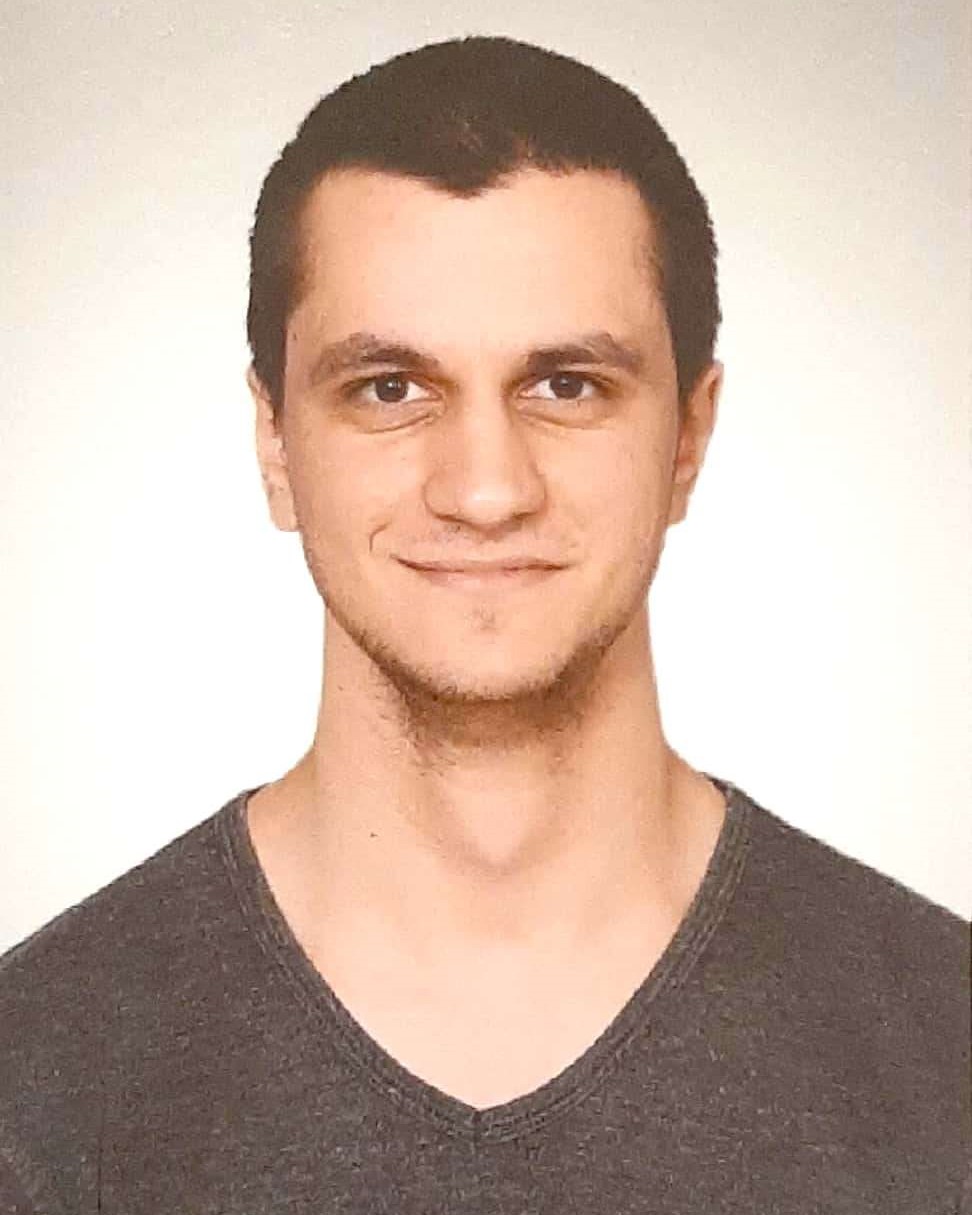}}
\end{minipage}
\hspace{0.5cm}
\begin{minipage}[H]{13.5cm}
\vspace{-1.4cm}\textbf{Okyanus Oral} received the B.Sc. degree from the Department of Electrical and Electronics Engineering from Middle East Technical University (METU), Ankara, Turkey, in 2021. He has been pursuing his M.Sc. degree in the same department since 2021 and has been a Teaching/Research Assistant since 2022. His research interests include inverse problems, computational imaging, optimization, and deep learning.
\end{minipage}
\bigskip
\bigskip
\bigskip
\bigskip
\bigskip
\bigskip

\begin{minipage}[H]{2.5cm}
\vspace{-2.4cm}
{\includegraphics[width=1in,height=1.25in,clip,keepaspectratio]{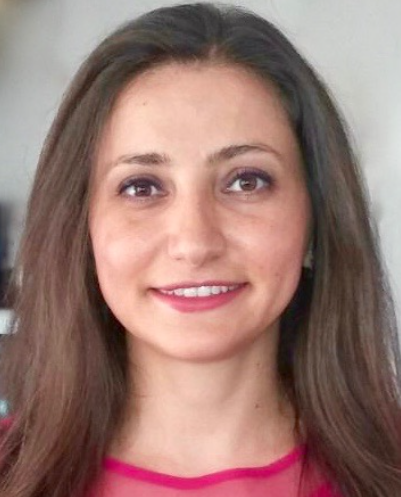}}\end{minipage}
\hspace{0.5cm}
\begin{minipage}[H]{13.5cm}
\vspace{-1.4cm}\textbf{Figen S. Oktem} (M’08) received the B.S. and M.S. degrees in electrical engineering from Bilkent University, Turkey, in 2007 and 2009, respectively, and the Ph.D. degree in electrical and computer engineering from the University of Illinois at Urbana-Champaign (UIUC), USA, in 2014. At UIUC, she was selected to the “List of Teachers Ranked as  Excellent by Their Students”, and was a recipient of NASA Earth and  Space Science Fellowship and Professor Kung Chie Yeh Endowed  Fellowship. She was then a Postdoctoral Research Associate with the NASA Goddard Space Flight Center, where she worked on high-resolution spectral imaging. She is now an Associate Professor in the Department of Electrical and Electronics Engineering at Middle East Technical University (METU). Her research spans the areas of computational imaging, inverse problems, statistical signal processing, machine learning, compressed sensing, and optical information processing. She is a member of the IEEE and Optica. 
\end{minipage}

\newpage
\thispagestyle{plain}
\setcounter{page}{1}
\setcounter{section}{0}
\setcounter{figure}{0}

\begin{center}
\par\noindent\rule{\textwidth}{2pt}
\vspace{0.003cm}

\LARGE\sc Plug-and-Play Regularization on Magnitude with Deep\\ Priors for 3D Near-Field MIMO Imaging: Supplementary Material\vspace{0.002cm}
\par\noindent\rule{\textwidth}{2pt}
\end{center} 

\begin{center}
\normalfont Okyanus~Oral\,\orcidlink{0000-0001-5059-4351},~\textit{Graduate~Student~Member}, and Figen~S.~Oktem\,\orcidlink{0000-0002-7882-5120},~\textit{Member,~IEEE}
\end{center}

\blfootnote{This work is supported by the Scientific and Technological Research Council of Turkey  (TUBITAK) under grant 120E505.}
\blfootnote{O.~Oral, and F.~S.~Oktem are with the Department of Electrical Engineering, METU, Cankaya, Ankara 06800, Turkey (e-mail: ookyanus@metu.edu.tr, figeno@metu.edu.tr).}
\blfootnote{This work has been submitted to the IEEE for possible publication. Copyright may be transferred without notice, after which this version may no longer be accessible.}

\section{Denoising Performance}

\begin{figure}[H]
     \centering
    \begin{subfigure}[H]{0.66\textwidth}
         \centering
         \includegraphics[width=\textwidth]{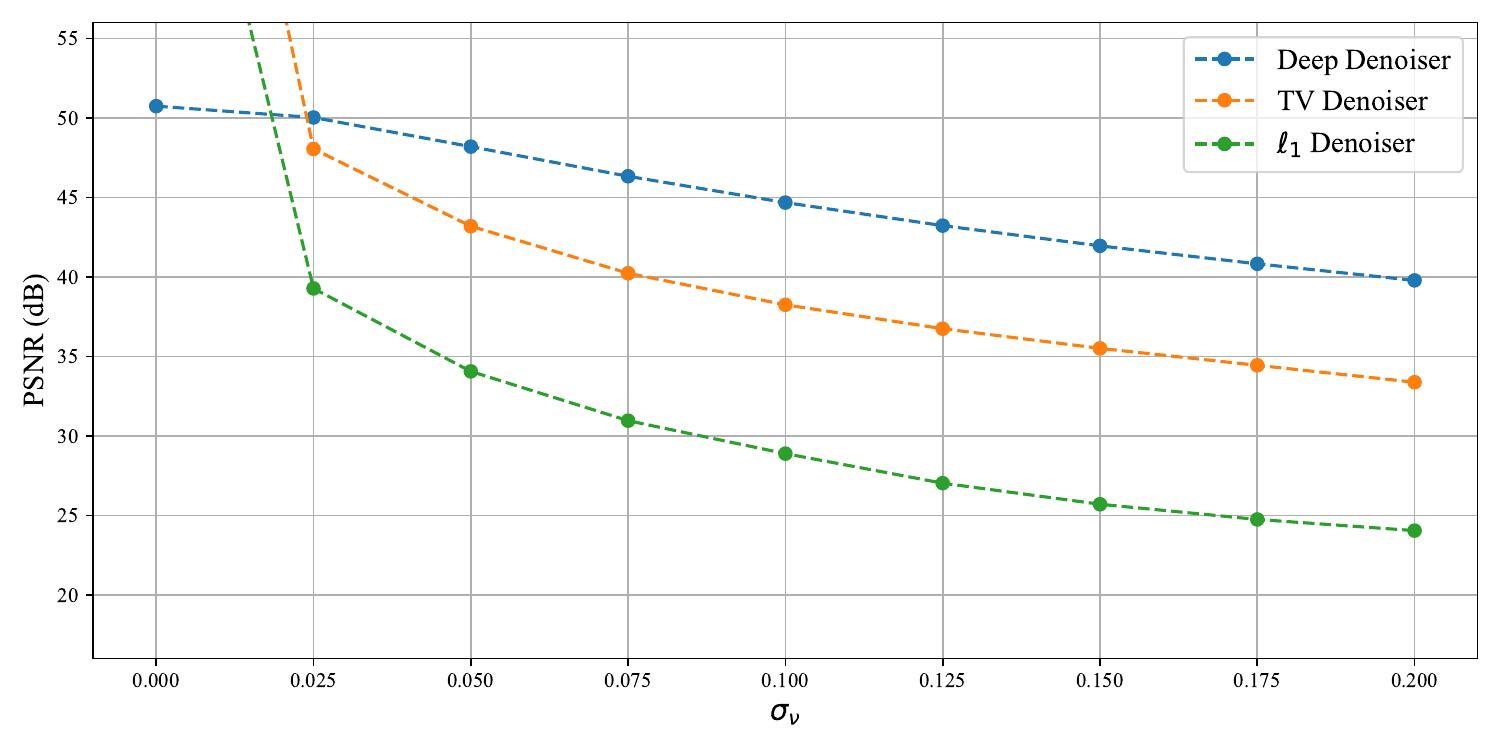}
         \caption{Average Denoising PSNR}
         \label{fig:DenoisingTest}
     \end{subfigure}
     \hfill
    \begin{subfigure}[H]{0.33\textwidth}
         \centering
         \includegraphics[width=\textwidth]{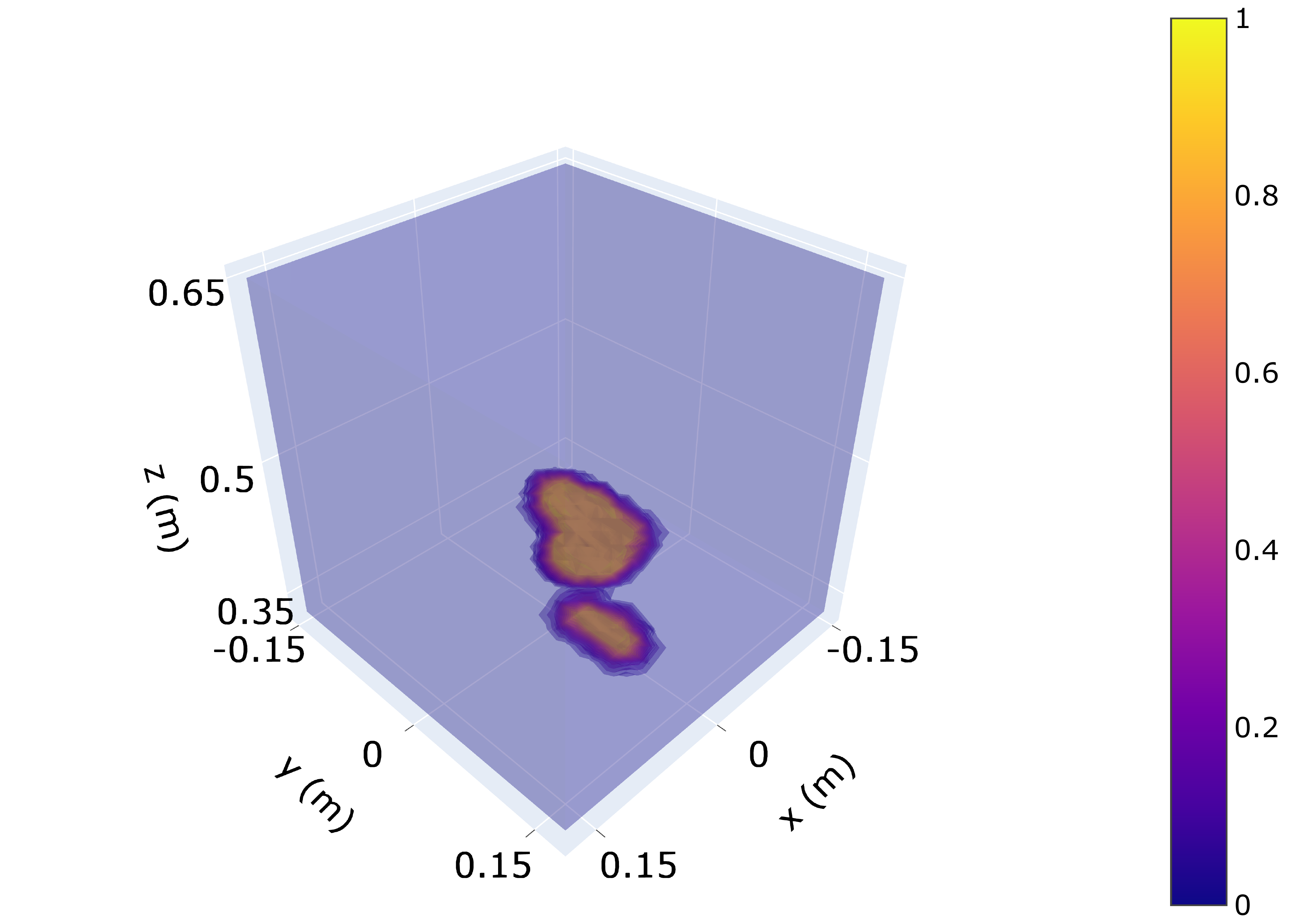}
         \caption{Ground Truth}
     \end{subfigure}
     \\
     \begin{subfigure}[H]{0.23\textwidth}
         \centering
         \includegraphics[width=\textwidth]{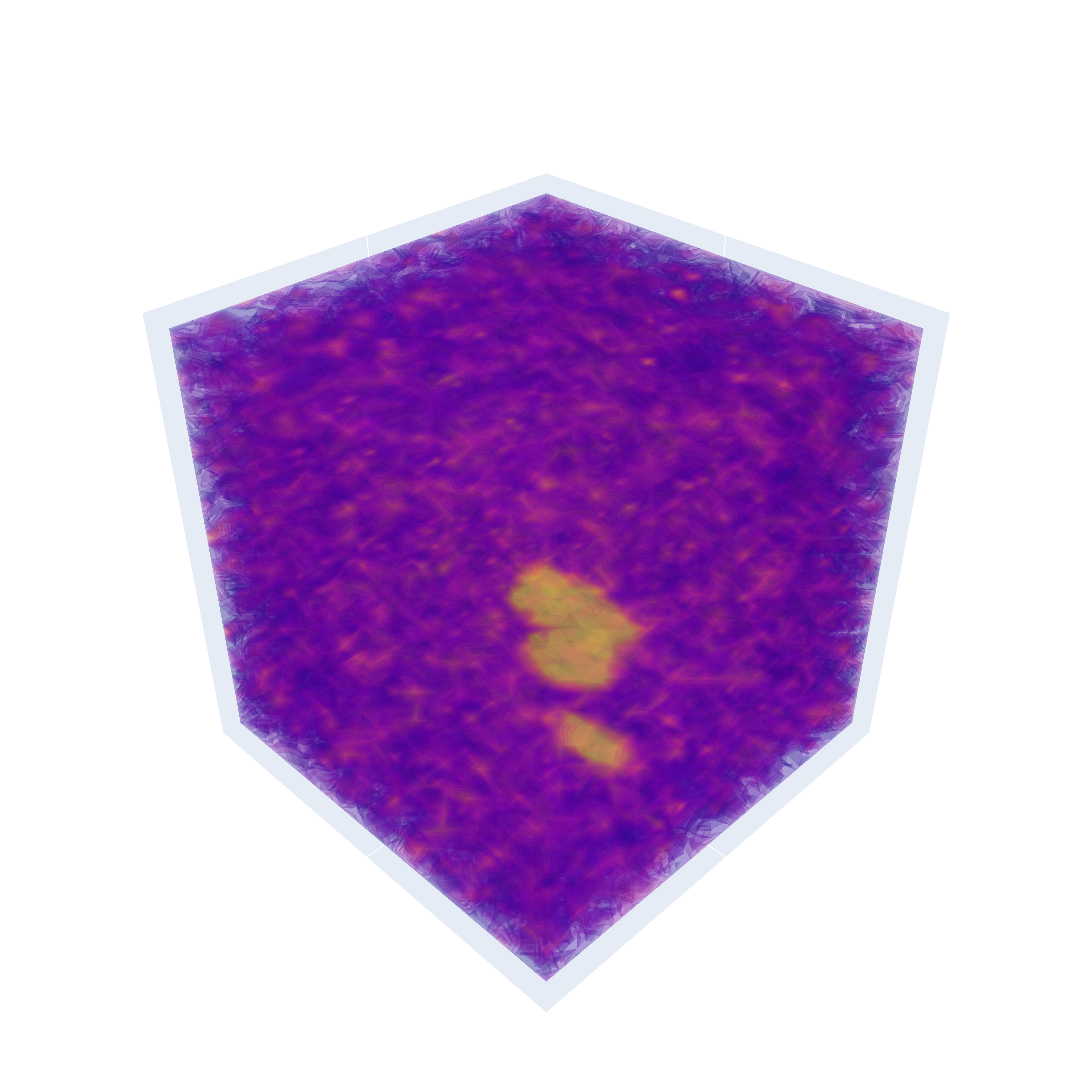}
         \caption{Input Magnitudes}
     \end{subfigure}
     \hfill
     \begin{subfigure}[H]{0.23\textwidth}
         \centering
         \includegraphics[width=\textwidth]{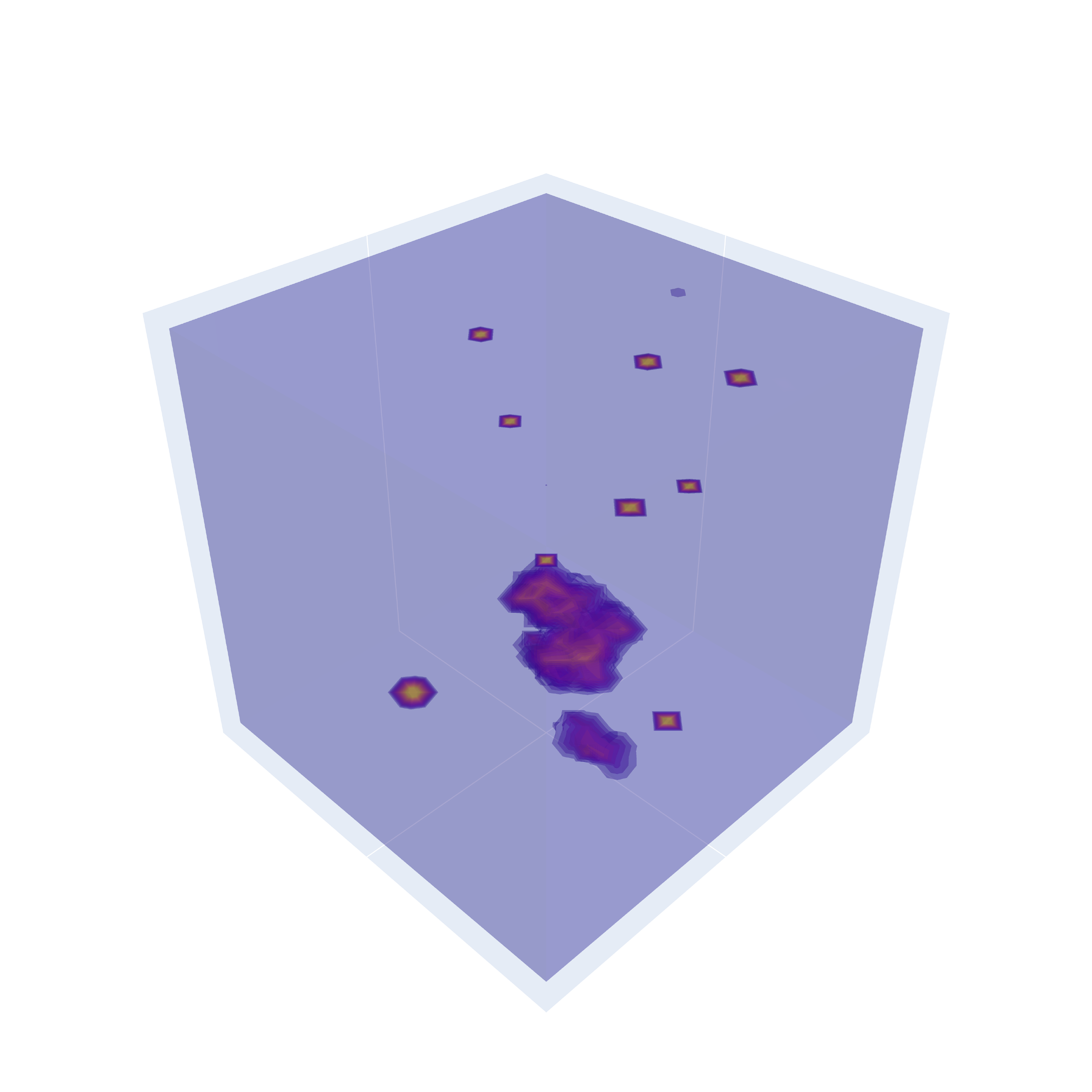}
         \caption{$\ell_1$: 23.66 dB}
     \end{subfigure}
     \hfill
     \begin{subfigure}[H]{0.23\textwidth}
         \centering
         \includegraphics[width=\textwidth]{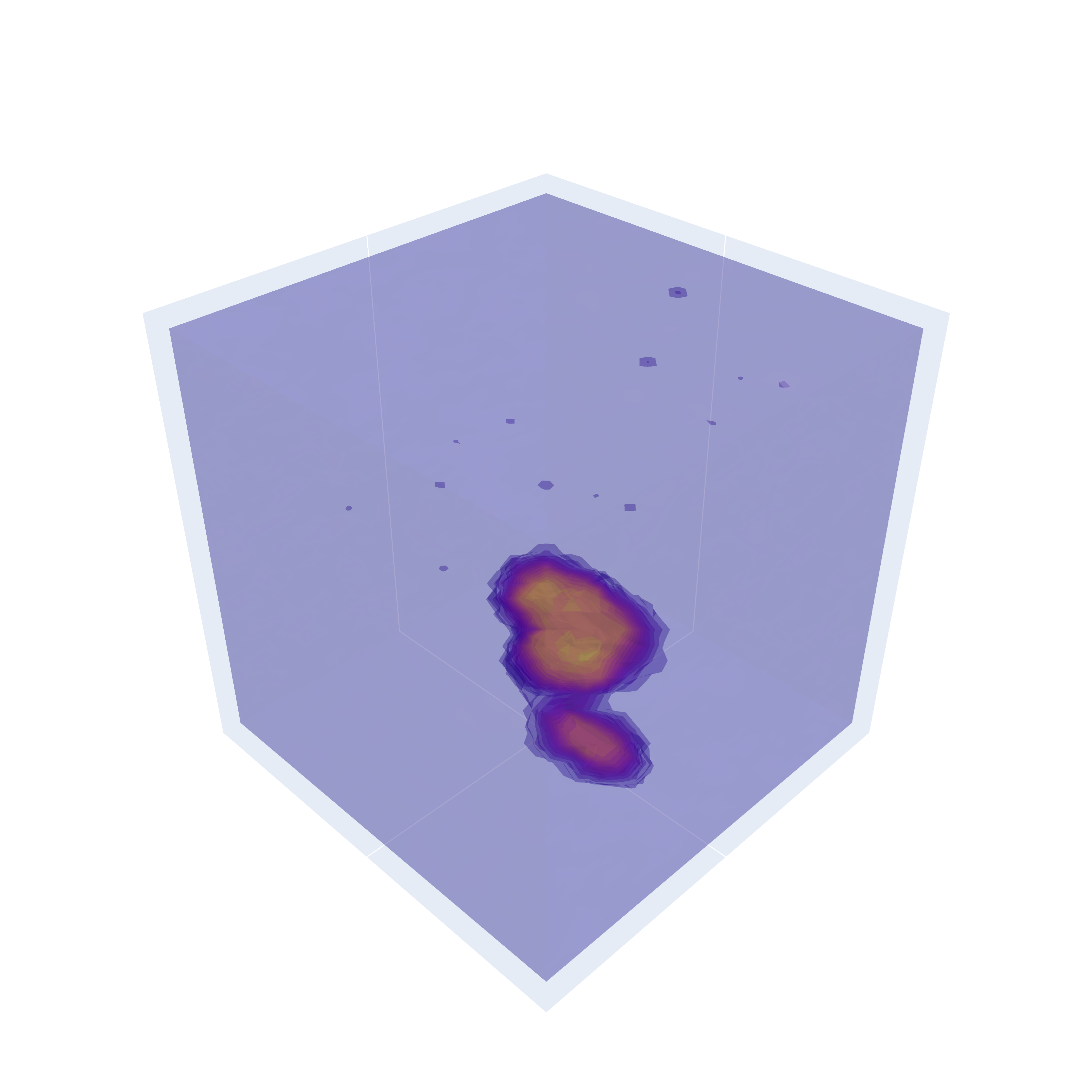}
         \caption{TV: 33.45 dB}
     \end{subfigure}
     \hfill
    \begin{subfigure}[H]{0.23\textwidth}
         \centering
         \includegraphics[width=\textwidth]{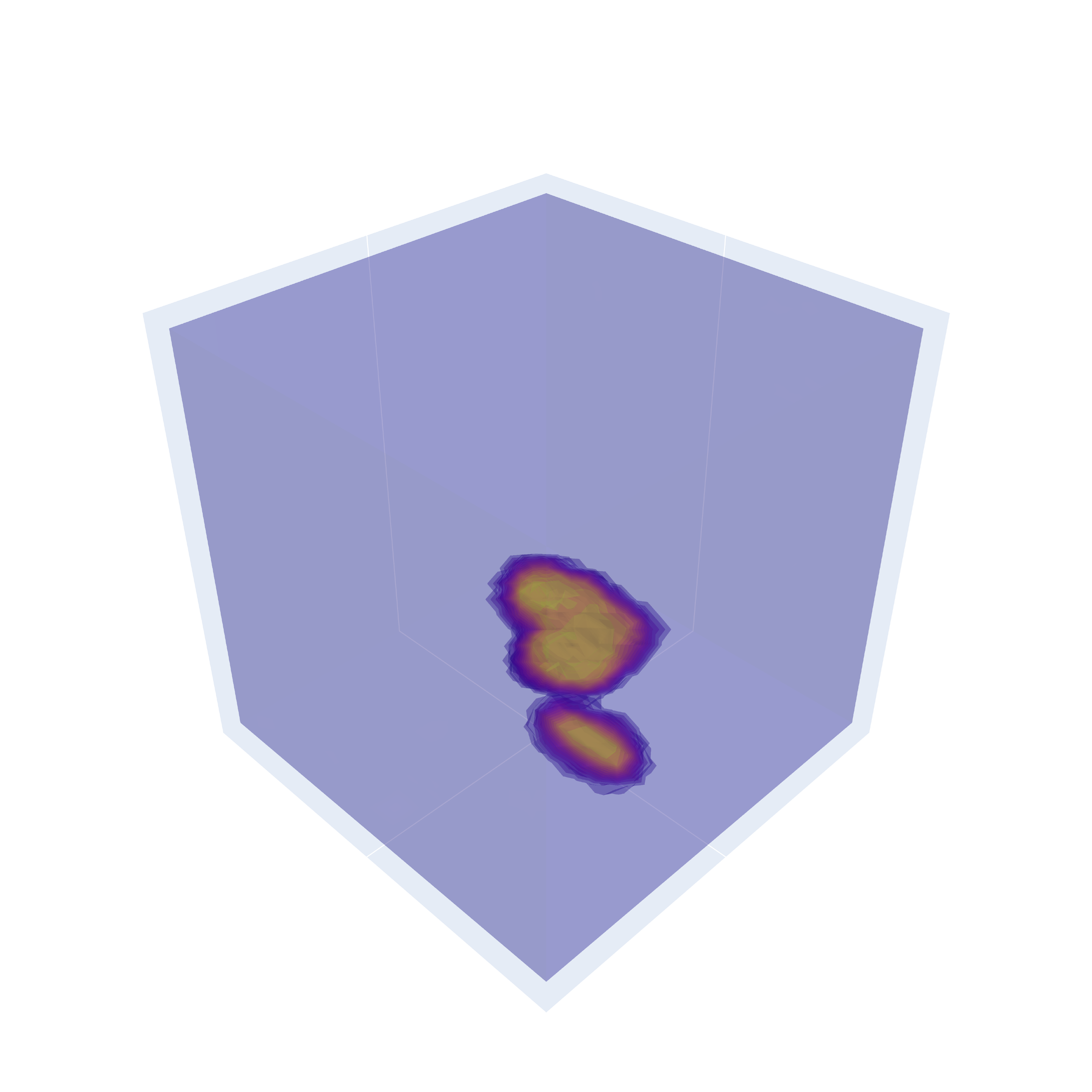}
         \caption{DNN: 39.85dB}
     \end{subfigure}

        \caption{Denoising performance of different methods; (a) average test PSNR with respect to noise standard deviation $\sigma_\nu$, (b) ground truth magnitudes of the sample test image, (c) noisy input magnitudes at $\sigma_\nu=0.2$, (d)-(f) denoised outputs corresponding to $\ell_1$, $TV$ and deep-prior based denoising and the respective PSNRs (dB).}
    \label{fig:Denoising}
\end{figure}

 Here we present the denoising performance of the trained DNN in comparison with the other denoising approaches ($\ell_1$ and TV regularization). %for denoising the magnitude of the  datacubes in the test dataset. 
The average PSNR is computed using 100 test images for different values of noise standard deviation $\sigma_\nu$ and provided in Fig.~\ref{fig:Denoising}a. Sample denoised magnitudes are also shown in Fig.~\ref{fig:Denoising}b-f. As seen, the deep denoiser significantly outperforms other methods.

Data-driven DNN-based denoisers are currently the best choice for plug-and-play regularization because DNNs provides state-of-the-art performance for the denoising problem as demonstrated in various works in the literature~[1]. In contrast to the existing analytical (hand-crafted) denoisers such as those based on $l_1$ and TV regularization, DNN-based denoisers are data-adaptive denoisers that learn how to remove the noise for the data of interest. Since the parameters of the deep denoiser are optimized based on the training data, prior information about the target images is learned. On the other hand, TV and $\ell_1$ regularization functions are hand-crafted and correspond to much simpler priors. 

\section{Reconstruction at a Finer Spatial Resolution}

\begin{figure}[H]
     \centering
      \begin{subfigure}[H]{0.31\textwidth}
         \centering
         \includegraphics[width=\textwidth]{figures/revolver10.png}
         \caption{Photograph of the toy revolver [2],[3].}
         \label{fig:DenoisingTest}
     \end{subfigure}
     \hfill
     \begin{subfigure}[H]{0.65\textwidth}
         \centering
         \includegraphics[scale=1]{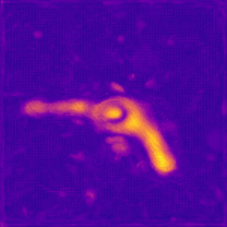}
         \includegraphics[scale=1]{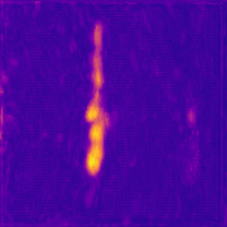}
         \includegraphics[scale=0.35]{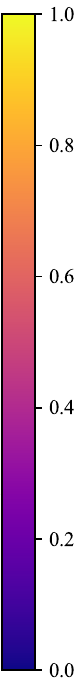}
         \caption{Reconstructed Image of size $151\times151\times151$; left, projection onto the $x-y$ plane, right projection onto the $y-z$ plane.}
         \label{fig:DenoisingTest}
     \end{subfigure}
     
    \caption{Imaged revolver and its reconstructions using 11 frequency steps with experimental data. Images have a 2mm resolution.}
    \label{fig:Splitting}
\end{figure}

Here, we analyze the performance of our approach at a finer spatial resolution. We tested our approach for a datacube of size $151\times151\times151$ within the same physical space (with 2mm resolution) using 11 frequency steps. Since 3D rendering becomes difficult at this grid size, we provide the maximum projections of the obtained reconstruction in Fig.~\ref{fig:Splitting}. As seen in this figure, we do not observe any splitting behavior with increased spatial resolution.

We expect the approach to provide similar performance at finer resolutions as long as compression level (data availability) kept similar and finer resolution used also for the training dataset. As the spatial resolution of the test object digresses away from the training dataset's resolution, the performance can inevitably be affected. However, we still observe good performance for both the presented result in the manuscript with grid size $61\times61\times61$ and the result presented here with the higher grid size $151\times151\times151$ although grid size of $25\times25\times49$ has been used for the training dataset. 

\section*{References}
\begin{minipage}[H]{0.1\textwidth}
    \hspace{.14cm}[1]
    \vspace{0.4cm}
\end{minipage}
\hspace{-1cm}
\begin{minipage}[H]{0.95\textwidth}
 G. Ongie, A. Jalal, C. A. M. R. G. Baraniuk, A. G. Dimakis, and R. Willett, “Deep learning techniques for inverse problems
in imaging,” \textit{IEEE Journal on Selected Areas in Information Theory}, vol. 1, no. 1, pp. 39–56, 2020.
\end{minipage}

\begin{minipage}[H]{0.1\textwidth}
    \hspace{.14cm}[2]
    \vspace{0.71cm}
\end{minipage}
\hspace{-1cm}
\begin{minipage}[H]{0.95\textwidth}J. Wang, P. Aubry, and A. Yarovoy, “3-D short-range imaging with irregular MIMO arrays using NUFFT-based range
migration algorithm,” \textit{IEEE Transactions on Geoscience and Remote Sensing}, vol. 58, no. 7, pp. 4730–4742, 2020.
\end{minipage}

\begin{minipage}[H]{0.1\textwidth}
    \hspace{.14cm}[3]
    \vspace{0.42cm}
\end{minipage}
\hspace{-1cm}
\begin{minipage}[H]{0.95\textwidth}J. Wang, “EM data acquired with irregular planar MIMO arrays,” 2020. [Online]. Available: \hyperlink{https://dx.doi.org/10.21227/
src2-0y50}{https://dx.doi.org/10.21227/
src2-0y50}
\end{minipage}

\end{document}